\documentclass[twocolumn,preprintnumbers,superscriptaddress,aps,nofootinbib]{revtex4-2}
\usepackage{amsmath,amssymb,mathtools}
\usepackage{graphicx}
\usepackage{booktabs,cellspace}
\usepackage{color}
\usepackage{hyperref}
\usepackage{xspace}
\usepackage[normalem]{ulem}
\usepackage{soul}
\usepackage{mathrsfs}

\makeatletter 
\renewcommand\onecolumngrid{
\do@columngrid{one}{\@ne}%
\def\set@footnotewidth{\onecolumngrid}
\def\footnoterule{\kern-6pt\hrule width 1.5in\kern6pt}%
}
\makeatother

\newcommand{\itilde}{{\tilde \imath}}

\newcommand{\ktilde}{{\tilde k}}
\newcommand{\betaps}{\beta_\text{\textsc{ps}}}
\newcommand{\betaobs}{\beta_\text{obs}}

\newcommand{\as}{\ensuremath{\alpha_s}\xspace}
\newcommand{\aeff}{\ensuremath{\alpha_\text{eff}}\xspace}

\newcommand{\drift}[1]{\langle \Delta_{#1} \rangle}

\newcommand{\GeV}{\ensuremath{\,\mathrm{GeV}}\xspace}

\newcommand{\ps}{\text{\textsc{ps}}}
\newcommand{\resum}{\text{resum}}

\newcommand{\LL}{\text{\textsc{ll}}}
\newcommand{\NLL}{\text{\textsc{nll}}}
\newcommand{\NNLL}{\text{\textsc{nnll}}}

\newcommand{\MSbar}{\overline{\textsc{ms}}}

\newcommand{\nlo}{\text{\textsc{nlo}}}
\newcommand{\nll}{\text{\textsc{nll}}}
\newcommand{\nnll}{\text{\textsc{nnll}}}

\newcommand{\order}[1]{\mathcal{O}\left(#1\right)}

\newcommand{\logbook}[2]{}

\newcommand{\cmw}{\text{\textsc{cmw}}}
\newcommand{\cB}{\mathcal{B}}
\newcommand{\cV}{\mathcal{V}}
\newcommand{\cR}{\mathcal{R}}
\newcommand{\cF}{\mathcal{F}}
\newcommand{\cP}{\mathcal{P}}

\definecolor{darkgreen}{rgb}{0,0.4,0}
\definecolor{amaranth}{rgb}{0.9,0.17,0.31}
\definecolor{grey}{rgb}{0.5,0.5,0.5}
\definecolor{orange}{rgb}{0.9,0.5,0.0}
\definecolor{lightblue}{rgb}{0.0,0.5,1.0}
\usepackage[dvipsnames]{xcolor}

\newcommand{\OXaff}{Rudolf Peierls Centre for Theoretical Physics,
  Clarendon Laboratory, Parks Road, Oxford OX1 3PU, UK} 
\newcommand{\CERNaff}{CERN, Theoretical Physics Department, CH-1211 Geneva 23, Switzerland}
\newcommand{\ASCaff}{All Souls College, Oxford OX1 4AL, UK}
\newcommand{\UCLaff}{Department of Physics and Astronomy, University College London, London, WC1E 6BT, UK}
\newcommand{\IPhTAff}{IPhT, Universit\'{e} Paris-Saclay, CNRS UMR 3681,
  CEA Saclay, F-91191 Gif-sur-Yvette, France}
\newcommand{\MonashAff}{School of Physics and Astronomy, Monash
  University, Wellington Rd, Clayton VIC-3800, Australia}
\newcommand{\ManAff}{Department of Physics \& Astronomy, University of
  Manchester, Manchester M13 9PL, United Kingdom}
\newcommand{\NIKHEFAff}{Nikhef, Theory Group, Science Park 105, 1098 XG, Amsterdam, The Netherlands}
\begin{document}

\title{
  A new standard for the logarithmic accuracy of parton showers
}

\preprint{CERN-TH-2024-057, OUTP-24-03P}

\author{Melissa van Beekveld}     \affiliation{\NIKHEFAff}
\author{Mrinal Dasgupta}          \affiliation{\ManAff} %
\author{Basem Kamal El-Menoufi}   \affiliation{\MonashAff} %
\author{Silvia~Ferrario~Ravasio}  \affiliation{\CERNaff}  %
\author{Keith Hamilton}           \affiliation{\UCLaff}  %
\author{Jack Helliwell}           \affiliation{\OXaff}%
\author{Alexander Karlberg}       \affiliation{\CERNaff}  %
\author{Pier Francesco Monni}     \affiliation{\CERNaff}  %
\author{Gavin P.\ Salam}          \affiliation{\OXaff}\affiliation{\ASCaff}%
\author{Ludovic Scyboz}           \affiliation{\MonashAff}  %
\author{Alba Soto-Ontoso}         \affiliation{\CERNaff}%
\author{Gregory Soyez}            \affiliation{\IPhTAff}%

\begin{abstract}
  We report on a major milestone in the construction of
  logarithmically accurate final-state parton showers, achieving
  next-to-next-to-leading-logarithmic (NNLL) accuracy for the wide
  class of observables known as event shapes.
  The key to this advance lies in the identification of the relation
  between critical NNLL analytic resummation ingredients and their
  parton-shower counterparts.
  Our analytic discussion is supplemented with numerical tests of the
  logarithmic accuracy of three shower variants for more than a 
  dozen distinct event-shape observables in $Z\to q\bar q$ and
  Higgs\,$\to gg$ decays.
  The NNLL terms are phenomenologically sizeable, as illustrated in
  comparisons to data. 
\end{abstract}

\maketitle

\logbook{}{Detailed info to be found in logbook/2023-07-21-NNLL-ee/}

Parton showers are essential tools for predicting QCD physics
at colliders across a wide range of momenta from the TeV down
to the GeV
regime~\cite{10.21468/SciPostPhys.16.5.130,Sherpa:2019gpd,Bierlich:2022pfr,Bewick:2023tfi}.
In the presence of such disparate momenta, the perturbative
expansions of quantum field theories have coefficients enhanced by
large logarithms of the ratios of momentum scales.
One way of viewing parton showers is as automated and
immensely flexible tools for resumming those logarithms, thus
correctly reproducing the corresponding physics.

The accuracy of resummations is usually classified based on terms with
the greatest logarithmic power at each order in the strong coupling
(leading logarithms or LL), and then towers of terms with 
subleading powers of logarithms at each order in the coupling (next-to-leading
logarithms or NLL, NNLL, etc.).
Higher logarithmic accuracy for parton showers should make them
considerably more powerful tools for analysing and interpreting
experimental data at CERN's Large Hadron Collider and potential future
colliders.
The past years have seen major breakthroughs in advancing the
logarithmic accuracy of parton showers, with
several groups taking colour-dipole showers from LL to
NLL~\cite{Dasgupta:2020fwr,Hamilton:2020rcu,Karlberg:2021kwr,Hamilton:2021dyz,vanBeekveld:2022zhl,vanBeekveld:2022ukn,vanBeekveld:2023chs,Forshaw:2020wrq,Nagy:2020dvz,Nagy:2020rmk,Herren:2022jej,Assi:2023rbu,Preuss:2024vyu,Hoche:2024dee}. 
There has also been extensive work on incorporating higher-order
splitting kernels into
showers~\cite{Jadach:2011kc,Hartgring:2013jma,Jadach:2013dfd,Jadach:2016zgk,Li:2016yez,Hoche:2017iem,Hoche:2017hno,Dulat:2018vuy,Campbell:2021svd,Gellersen:2021eci,FerrarioRavasio:2023kyg}
and understanding the structure of subleading-colour
corrections, see e.g.\
Refs.~\cite{Nagy:2015hwa,Nagy:2019pjp,Nagy:2019rwb,DeAngelis:2020rvq,Holguin:2020joq,Hoche:2020pxj,Hamilton:2020rcu,Hatta:2013iba,Hagiwara:2015bia,Hatta:2020wre,Becher:2021zkk,Becher:2023mtx,Boer:2024hzh}.

Here, for the first time, we show how to construct parton showers with
NNLL accuracy for the broad class of event-shape observables at lepton colliders,
like the well-known Thrust~\cite{Brandt:1964sa,Farhi:1977sg} (see e.g.\
Refs.~\cite{%
  deFlorian:2004mp,
  Becher:2008cf,
  Abbate:2010xh,
  Chien:2010kc,
  Monni:2011gb,    
  Becher:2012qc,   
  Hoang:2014wka,   
  Banfi:2014sua,   
  Banfi:2016zlc,   
  Frye:2016okc,    
  Frye:2016aiz,    
  Tulipant:2017ybb,
  Moult:2018jzp,   
  Bell:2018gce,    
  Banfi:2018mcq,   
  Procura:2018zpn, 
  Arpino:2019ozn,  
  Bauer:2020npd,   
  Kardos:2020gty,  
  Anderle:2020mxj, 
  Dasgupta:2022fim,
  Duhr:2022yyp}     
for calculations at NNLL and beyond).  
This is achieved by developing a novel framework that unifies several 
recent developments, on (a)
the inclusive structure of soft-collinear gluon
emission~\cite{Banfi:2018mcq,Catani:2019rvy} up to third order in the
strong coupling $\as$;
(b) the inclusive pattern of energetic (``hard'')  collinear
radiation up to order
$\as^2$~\cite{Dasgupta:2021hbh,vanBeekveld:2023lsa};
and (c) the incorporation of soft radiation fully differentially up to
order $\as^2$ in parton showers, ensuring correct generation of any
number of well-separated pairs of soft
emissions~\cite{FerrarioRavasio:2023kyg}.
These are all NNLL after integration of the respective (a) double and
(b,c) single logarithmic phase spaces.

\begin{figure}
  \centering
  \includegraphics[width=\linewidth]{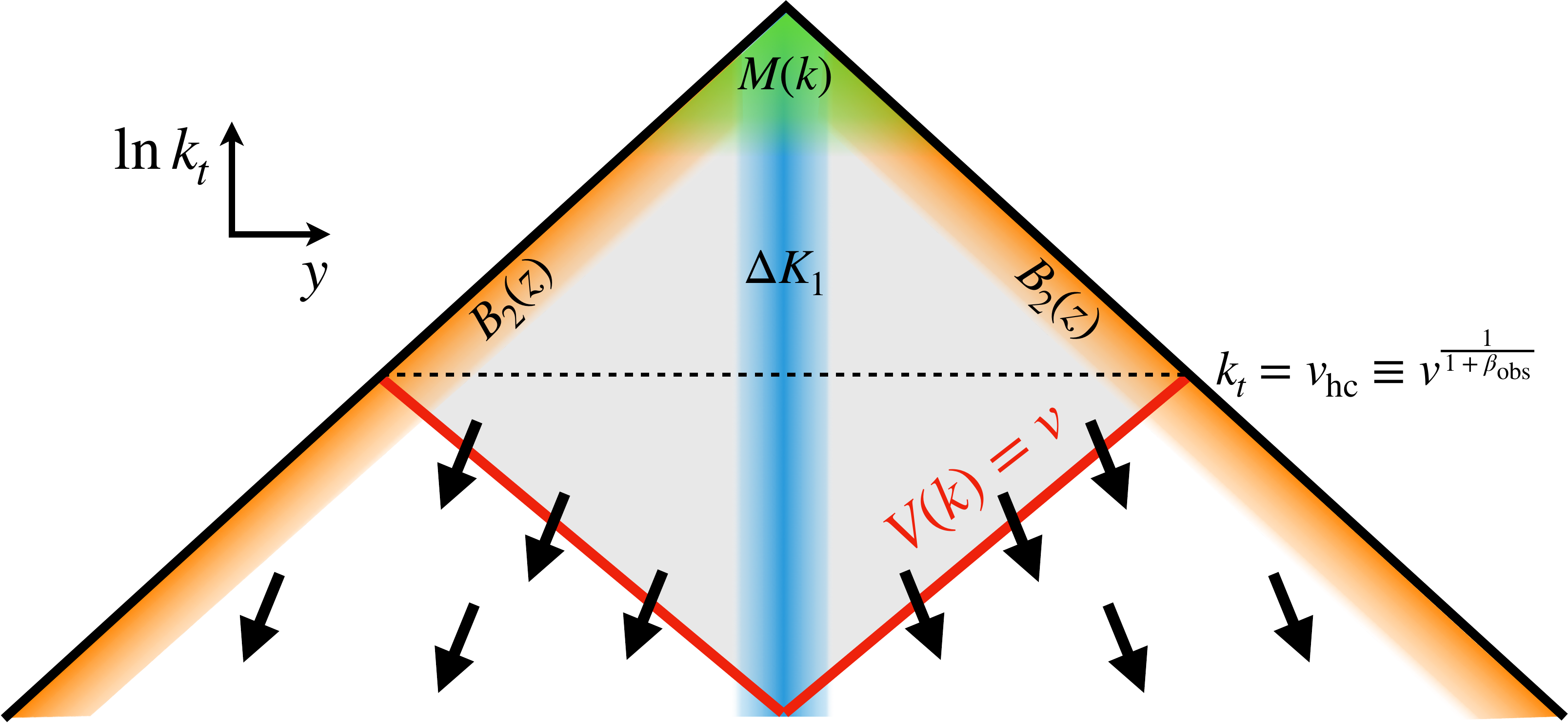}
  \caption{Schematic representation of the Lund plane~\cite{Andersson:1988gp}.
    A constraint on an event shape that scales as $k_t e^{-\betaobs
      |y|}$ implies that shower emissions above the red line are
    mostly vetoed. 
    At NNLL, one mechanism that modifies this
      constraint is that
    subsequent branching may cause the effective transverse momentum
    or rapidity to shift, as represented by the arrows.
  }
  \label{fig:drifts}
\end{figure}

We will focus the discussion on the $e^+e^- \to Z\to q\bar q$ process,
with the understanding that the same arguments apply also to
$H\to gg$.
Each event has a set of
emissions with momenta $\{k_i\}$ and we work in units where the
centre-of-mass energy $Q\equiv1$.
We will examine the probability $\Sigma(v)$ that some global event
shape, $V(\{k_i\})$, has a value $V(\{k_i\})<v$.
Event-shape observables have the property~\cite{Banfi:2004yd}
that for a single soft and collinear emission $k$,
$V(k) \propto k_t e^{-\betaobs |y|}$, where $k_t$ ($y$) is the
transverse momentum (rapidity) of $k$ with respect to the Born event
direction and $\betaobs$ depends on the specific observable, e.g.\
$\betaobs=1$ for Thrust.
Whether considering analytic resummation or a parton shower, for
$v \ll 1$ we have
\begin{equation}
  \label{eq:Sigma-main}
  \Sigma(v) \!=\! \mathcal{F}
  \exp
    \!\left[\!
    -4 \!\!\int \! \frac{dk_t}{k_t}\!\!
    \int_{k_t}^1 \!\!
    \!\!dz P_{\!gq}(z) M(k) \frac{\aeff}{2\pi}
    \Theta(V(k) \!>\! v)
  \right]\!\!,
\end{equation}
with $P_{gq}(z)=C_F \frac{1+(1-z)^2}{z}$ and $M(k)$ a function that
accounts for next-to-leading order matching, with $M(k) \to 1$ for
$k_t \to 0$.
The exponential is a Sudakov form factor, encoding the suppression of
emissions with $V(k)>v$, cf.\ the grey region of
Fig.~\ref{fig:drifts}.
It brings the LL contributions to $\ln \Sigma$, terms $\as^n L^{n+1}$
with $L = \ln v$, as well as NLL ($\as^nL^n$), NNLL ($\as^nL^{n-1}$),
etc., contributions. 
The function $\cF$ accounts~\cite{Banfi:2004yd} for the difference
between the actual condition $V(\{k_i\})<v$ and the simplified
single-emission boundary $V(k)<v$ that is used in the Sudakov.
It starts at NLL.

In Eq.~(\ref{eq:Sigma-main}), the effective coupling, $\aeff$, can be
understood as the intensity of gluon emission, inclusive over possible
subsequent branchings of that emission and corresponding virtual
corrections.
We write it as
\begin{equation}
  \label{eq:alpha-eff}
  \aeff =
  \as\!
  \left[1 \!+
    \frac{\as}{2\pi}
    \left(
      K_1 \!+\! \Delta K_1(y) \!+\! B_2(z)
    \right)
    + \frac{\as^2}{4\pi^2} K_2
  \right]\!\!,
\end{equation}
with $\as \equiv \as^{\MSbar}(k_t)$ and here the rapidity $y = \ln
z/k_t$.
$K_1=(\frac{67}{18}-\frac{\pi^2}{6}) C_A - \frac{10}{9}
n_fT_R$ (often called $K_\cmw$) \cite{Catani:1990rr} is required for NLL
accuracy and the remaining terms for NNLL.
$\Delta K_1(y)$ is zero in the resummation literature, non-zero at
central rapidities for certain showers, and vanishes for
$y \to \infty$~\cite{FerrarioRavasio:2023kyg};
$B_2(z)$ affects the hard-collinear region and tends to zero in the
soft limit, $z\to0$.
In analytic resummation it is generally included as a constant
multiplying $\delta(1-z)$~\cite{deFlorian:2004mp}.
It has been calculated in specific resummation schemes in
Refs.~\cite{Dasgupta:2021hbh,vanBeekveld:2023lsa}, but is not yet
known for the showers that we consider, which also do not yet include
the relevant triple-collinear dynamics.
At NNLL, $K_2$ is relevant in the whole soft-collinear region and also
so far calculated only for analytic
resummation~\cite{Banfi:2018mcq,Catani:2019rvy}.
Through shower unitarity, $\aeff$ is relevant also in the Sudakov-veto
region of Fig.~\ref{fig:drifts}, i.e.\ the region above the red line,
even though that region contains no emissions or subsequent branching.

It is straightforward to see from Eq.~(\ref{eq:Sigma-main}) that terms
up to $\as^n L^{n-1}$ in $\ln \Sigma(v)$ depend only on the integrals
of $\Delta K_1(y)$ and $B_2(z)$,
\begin{equation}
  \label{eq:DeltaKint-B2int-defs}
  \Delta K_1^\text{int} \equiv \!\int_{-\infty}^\infty \!\!\!\!\! dy\,
  \Delta K_1(y)\,,\,\,
  B_2^\text{int} \equiv\! \int_0^1 dz \frac{P_{gq}(z)}{2C_F} B_2(z).
\end{equation}
One of the key observations of this Letter is that as long as a
parton shower correctly generates double-soft emissions in the
soft-collinear region (as required for NNLL handling of event shapes'
sensitivity to multiple emissions),
it is then possible and sufficient to identify NNLL relations
between the $\Delta K_1^\text{int}$, $B_2^\text{int}$ and $K_2$ in event-shape resummation and the corresponding constants needed
for a parton shower.
This holds even if the shower does not reproduce the full
relevant physics at second order in the large-angle and hard-collinear
regions and at third order in the soft-collinear region.
This is in analogy with the fact that including the correct
$K_1$ constant is sufficient to obtain NLL accuracy even
without the real double-soft contribution.

In the next few paragraphs we will identify the relations between
individual resummation and shower ingredients (neglecting terms beyond
NNLL), and then show how they combine to achieve overall NNLL
shower accuracy.
Let us start by recalling how the $\order{\as^2}$
terms of Eq.~(\ref{eq:alpha-eff}) come about.
Consider a Born squared matrix element, $\cB_\itilde$, for producing
a gluon $\itilde$ (multiplied by $\as/2\pi$).
Schematically the $\order{\as^2}$ terms involve the single-emission
virtual correction $\cV_{\itilde}$ and an integral over a real
$\itilde \to ij$ branching phase space and matrix element,
$d\Phi_{ij|\itilde} \cR_{ij}$ (both multiply $(\as/2\pi)^2$).
The key difference between a resummation calculation and a parton
shower lies in the phase-space mapping that is encoded in
$d\Phi_{ij|\itilde}$. 
For example, in many resummation calculations $g_\itilde\to q_i\bar q_j$
splitting implicitly conserves transverse momentum
${k}_{t,i+j} = {k}_{t\itilde}$ and rapidity $y_{i+j} = y_{\itilde}$
with respect to the particle that emitted $\itilde$~\cite{Banfi:2018mcq,Catani:2019rvy}.
A parton shower ($\ps$) will organise the phase space differently, and
in a way that does not conserve these kinematic quantities.
The difference can be represented as an effective drift in one or more
kinematic variables $x$ (e.g. $x \equiv \ln k_t$, $x\equiv y$) of
post- versus pre-branching kinematics.
The average drifts,
$\frac{\as}{2\pi} \drift{x}$, are represented as arrows in
Fig.~\ref{fig:drifts}.
For a soft-collinear (SC) gluon $k_\itilde$, they are independent
of the kinematics of $k_\itilde$. 
For the $C_F C_A$ and $C_F n_f$ colour channels they read  
\begin{equation}
  \label{eq:y-drift}
  \drift{x}
  =
  \lim_{\itilde \to \text{\textsc{sc}}}
  \frac{1}{\cB_\itilde} \int
  d\Phi_{ij|\itilde}^\ps \, \cR_{ij}
  \times
  (x_{i+j} - x_\itilde)\,.
\end{equation}
For the $C_F^2$ channel, one replaces $x_{i+j}$ with the
$x$ value of that of $i$ and
$j$ that corresponds to the larger shower ordering variable ($v_\ps =
k_t e^{-\betaps |y|}$).
Note that the sign of $\drift{y}$
depends on the sign of $y_\itilde$ (below, $y_\itilde>0$).

To understand the relation of $\drift{y}$ with
$\Delta K^\text{int}_1$, observe that a drift to large absolute
rapidities depletes radiation at central rapidities.
However the shower must correctly reproduce the total final amount of
radiation integrated over any rapidity window.
That can only be achieved with a value for $\Delta K^\text{int}_1$ that
generates just enough extra central radiation to compensate for the
drift-induced depletion.
Quantitatively, the following relation can be proven
(Ref.~\cite{NNLLsupplement}, \S\ref{sec:div-theorem})
\begin{equation}
  \label{eq:y-drift-constraint}
  \Delta K_1^\text{int,\ps}
  =
  2\drift{y}
  \,.
\end{equation}
As a numerical check, Table~\ref{tab:drift-values} shows the result of
$\Delta K_1^\text{int,\ps}$ as determined in
Ref.~\cite{FerrarioRavasio:2023kyg}, compared to
$\drift{y}$ as determined for this paper.
The results are given for three 
variants~\cite{Dasgupta:2020fwr,FerrarioRavasio:2023kyg} of the
PanGlobal shower.
The $\text{PG}_{\beta=0}$ and $\text{PG}_{\beta=0}^\text{sdf}$ showers
have $\betaps=0$ and differ in how the splitting probabilities are
assigned between the two dipole ends.
For all three variants, one observes good agreement between
$\Delta K_1^\text{int,\ps}$ and $2\drift{y}$.

\begin{table}
  \centering
  \begin{tabular}{ccr@.lr@.lr@.l}
    \toprule
    shower & colour
    & \multicolumn{2}{c}{$\frac1{4\pi}\Delta K_1^\text{int,\ps}$}
    & \multicolumn{2}{c}{$\frac{1}{2\pi}\drift{y}$}
    & \multicolumn{2}{c}{$\frac{1}{2\pi}\drift{\ln k_t}$}
    \\  \midrule    
                           & $C_F$    & \multicolumn{2}{c}{0} &  0&000018(39)& -1&953481(1)  \\
    PG$_{\beta=0}^\text{sdf}$ & $C_A$    & \multicolumn{2}{c}{0} &  0&000002(2) &  1&162602(2)  \\
                           & $n_fT_R$ & \multicolumn{2}{c}{0} & -0&0000003(3)& -0&1048049(3) \\ \midrule    
                           & $C_F$    & 0&04967(3)            &  0&049576(8) & -1&964624(6)  \\
    PG$_{\beta=0}$           & $C_A$    & 0&0323(5)             &  0&032107(4) &  1&174900(4)  \\
                           & $n_fT_R$ & 0&0040(1)             &  0&003962(1) & -0&104655(1)  \\ \midrule    
                           & $C_F$    & 1&6725(5)             &  1&672942(9) & -1&749920(5)  \\
    PG$_{\beta=\frac{1}{2}} $  & $C_A$    & 0&0172(11)            &  0&015303(5) &  1&172042(5)  \\
                           & $n_fT_R$ & 0&0535(2)             &  0&053476(1) & -0&094205(1)  \\
    \bottomrule
  \end{tabular}
  \caption{The $\Delta K_1^\text{int,\ps}$ and $\drift{y}$ and
    $\drift{\ln k_t}$ coefficients, including the relevant leading-$N_C$ colour factors
    ($2C_F=C_A=3$ and $n_f=5$).
    The errors on $\Delta K_1^\text{int,\ps}$ are systematic dominated and
    estimated only to within a factor of order $1$.
    Their impact on the NNLL tests below is an order of magnitude
    smaller than the accuracies of those tests.  }
  \label{tab:drift-values}
\end{table}
\logbook{}{
  GS-NOTE: uncertainties on the dK integration obtained by looking at
  unskewed dipoles, with linear and cubic integration, also including
  the uncertainty from the linear integration. Script is
  integrate-table.py in 2020-eeshower/analyses/double-soft-second-order/gregory-tests/res/dKCMW
  See also
  2020-eeshower/analyses/double-soft-second-order/gregory-tests/drift-field.cc,
  including the suggested command-line at the start. (+section 5 of
  the ee-NNLL logbook)
}

Turning to $B_2(z)$, the corresponding physics differentially in $z$
cannot yet be included in our showers, insofar as they lack
triple-collinear splitting.
However, we can use a constraint analogous to
Eq.~(\ref{eq:y-drift-constraint}) to determine the correct
$B_2^\text{int,\ps}$, starting from  the NLO $1\to 2$ calculations of
Refs.~\cite{Dasgupta:2021hbh,vanBeekveld:2023lsa}, which conserve
the light-cone momentum-fraction $z = m_t e^y=\sqrt{k_t^2+m^2}e^y$. 
Specifically (Ref.~\cite{NNLLsupplement}, \S\ref{sec:B2-coeffs}),
\begin{align}
  \label{eq:B2-relations-int}
  B_2^\text{int,\ps} \!\!=  B_2^\text{int,\textsc{nlo}} \!\!- \drift{\ln z},\,
\end{align}
with
\begin{equation}
  \label{eq:Delta-lnz}
  \drift{\ln z}
  = \drift{y} + \drift{\ln m_t}
  = \drift{y} + \drift{\ln k_t} - \frac{\beta_0\pi^2}{12}.
\end{equation}
The $\beta_0=(11 C_A-4n_f T_R)/6$ term arises from the relation
between the drifts in $m_t$ and $k_t$, which is
shower-independent~\cite{Banfi:2018mcq,Catani:2019rvy,NNLLsupplement}.\footnote{%
  In the $C_F$ channel one defines the drift from the single 
  parton with larger $k_t e^{-\betaps |y|}$, so $m_t= k_t$ and
  $\drift{\ln m_t}_{C_F}=\drift{\ln k_t}_{C_F}$.
}
Note that Eq.~(\ref{eq:B2-relations-int}) does not constrain the
functional form of $B_2^{\ps}(z)$.
To do so meaningfully would require a shower that incorporates
triple-collinear splitting functions.
For event-shape NNLL accuracy, any reasonable functional form for $B_2^{\ps}(z)$
is equally valid, as long as it has the correct integral.
We choose the simple ansatz $B_2^\ps(z) \propto z$, normalised so as
to satisfy Eq.~(\ref{eq:B2-relations-int}).
Note that in an analytical resummation, Eq.~(\ref{eq:Sigma-main})
would use
$B_2^\text{int,\resum} = B_2^\text{int,\textsc{nlo}} + \frac{\beta_0
  \pi^2}{12}$ (the $\frac{\beta_0 \pi^2}{12}$ term has the same origin as
in Eq.~(\ref{eq:Delta-lnz})). 

The next ingredient that we need is $K_2$, which, for resummations, has been calculated
in two schemes~\cite{Banfi:2018mcq,Catani:2019rvy}.
We adopt the scheme in which transverse momentum is conserved and
consider the amount of radiation in a (fixed-rapidity)
transverse-momentum window $k_{tb} < k_{t} < k_{ta}$, where $k_t$ is
the post-branching pair transverse momentum.
The total amount of radiation in the window should be the same in the
resummation and the shower.
In the shower specifically, one should account for the $\ln k_t$ drifts
through the lower and upper edges of the window, which involve $\as$
at scales $k_{tb}$ and $k_{ta}$ respectively.
Defining
$T_n(k_{tb}, k_{ta}) = \int_{k_{tb}}^{k_{ta}} \frac{dk_t}{k_t} \frac{\as^n(k_t)}{(2\pi)^n}$,
that yields the constraint
\begin{multline}
  \label{eq:K2-basis}
  K_2^{\resum} \, T_3(k_{tb}, k_{ta}) = K_2^{\ps}\, T_3(k_{tb}, k_{ta})
  \,+
  \\
  +
  \left(
       \frac{\as^2(k_{tb})}{4\pi^2}
    -  \frac{\as^2(k_{ta})}{4\pi^2} 
  \right) \drift{\ln k_t}
  \,,
\end{multline}
where the second line accounts for the drift contributions at the
edges.
Setting
\begin{equation}
  \label{eq:K2ps-relation}
  K_2^\ps = K_2^\text{resum} - 4\beta_0  \drift{\ln k_t},
\end{equation}
ensures Eq.~(\ref{eq:K2-basis}) is satisfied for all NNLL terms
$\as^{2+n} \ln^n k_{t1}/k_{t2}$, noting that for 1-loop running,
\begin{equation}
  \label{eq:T-swap}
  2n\beta_0 T_{n+1}(k_{tb}, k_{ta}) = [\as^n(k_{tb}) -
  \as^n(k_{ta})]/(2\pi)^n.
\end{equation}
The final element in the connection with analytic resummation is
$\cF$, which encodes the effect of emissions near the boundary
$V(k) \sim v$.
The shower generates this factor through the interplay
between real and virtual emission.
However $\cF^\ps$ differs from $\cF^\resum$ because of relative drifts across
the boundary (Ref.~\cite{NNLLsupplement}, \S\ref{sec:cF-derivation})
\begin{equation}
  \label{eq:cF-ps-v-resum}
  \frac{\cF^\ps}{\cF^\resum}
  = 1 + 8 C_F T_2(v,v_\text{hc})
    \left[
    \drift{y}
    -
    \frac1{\betaobs}\drift{\ln k_t}
  \right], 
\end{equation}
with $v_\text{hc} \equiv v^{\frac{1}{1+\betaobs}}$.
Concentrating on the right-hand half of the Lund plane in
Fig.~\ref{fig:drifts}, it encodes the fact that a positive $y$ drift
increases the number of events that pass the constraint
$V(\{k\}) < v$, because emissions to the left of the boundary move to
the right of the boundary, and vice-versa for a positive $\ln k_t$
drift.

We are now in a position to write the ratio of $\Sigma(v)$ in the
shower as compared to a resummation.
Assembling the contributions discussed above into
Eq.~(\ref{eq:Sigma-main}) yields
\begin{multline}
  \label{eq:adding-it-together}
  \frac{\Sigma^\ps(v)}{\Sigma^\resum(v)} - 1 =
  8C_F \bigg\{
    - \drift{y} T_2(v,1)
  \\
    + \left[
      \drift{y} + \drift{\ln k_t}
    \right] T_2\left(v_\text{hc},1\right)
    \\
    + \drift{\ln k_t}
    \left[
      \frac{1}{\betaobs} T_2\left(v,v_\text{hc} \right)
      -
      T_2\left(v_\text{hc}, 1 \right)
    \right]
    \\
    + \left[
    \drift{y}
    -
    \frac{1}{\betaobs}\drift{\ln k_t}
  \right]
   T_2\left(v, v_\text{hc}\right)
  \bigg\} = 0\,,
\end{multline}
up to NNLL.
The lines account, respectively, for the shower contributions to
$\Delta K_1$, $B_2$, $K_2$ (using Eq.~(\ref{eq:T-swap}) and then
trading rapidity and $k_t$ integrations) and $\cF$.
The fact that they add up to zero ensures shower NNLL accuracy for
arbitrary global event shapes.
The last line necessarily involves real double-soft emissions in the
soft-collinear region, thus tying the other three lines (which just
involve the Sudakov non-emission probability) to the shower's
double-soft emissions, as anticipated below
Eq.~(\ref{eq:DeltaKint-B2int-defs}).
%
%
The connection with the ARES NNLL
formalism~\cite{Banfi:2014sua,Banfi:2016zlc,Banfi:2018mcq} is discussed in
Ref.~\cite{NNLLsupplement}, \S\ref{sec:anl-nnll}.

\begin{figure}
  \centering
  \includegraphics[width=\linewidth]{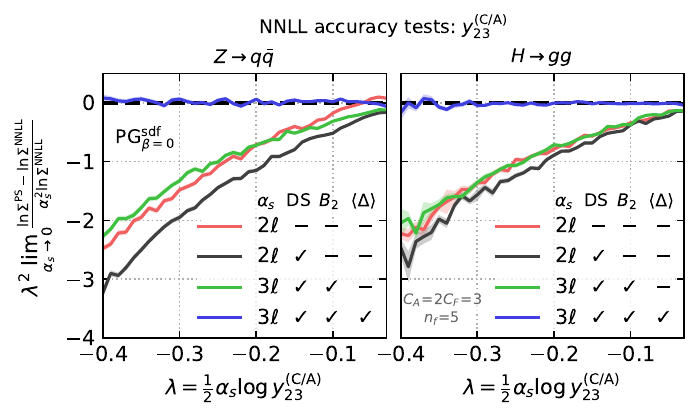}
  \caption{
    Test of NNLL accuracy of the PanGlobal
    (PG$_{\beta=0}^\text{sdf}$) shower for the cumulative distribution of the
    Cambridge $y_{23}$ resolution variable, compared to known
    results for $Z \to q\bar q$~\cite{Banfi:2016zlc} (left) and
    $H \to gg$~\cite{2024FutureResummations} (right).
    The curves show the difference relative to NNLL for various subsets of ingredients. Starting from
    the red curve, DS additionally includes double soft contributions
    and $2$-jet NLO matching; $3\ell$ includes 3-loop running of $\as$
    and the $K_2^\text{resum}$ term.
    $B_2$ in the legend refers only to its resummation part, $B_2^\text{int,NLO}$.
    Including all effects (blue line) gives a result that is
    consistent with zero, i.e.\ in agreement with NNLL.
  }
  \label{fig:log-dist-ratio}
\end{figure}

Besides the analytic proof, we also carry out a series of numerical
verifications of the NNLL accuracy of several parton showers with the
above elements, using a leading-colour limit $2C_F=C_A=3$.
These tests help provide confidence both in the overall picture and in
our specific implementation for final-state showers.
Fig.~\ref{fig:log-dist-ratio} shows a suitably normalised logarithm of
the ratio of the cumulative shower and resummed cross sections, for a
specific observable, the two-to-three jet resolution parameter, $y_{23}$, for the
Cambridge jet algorithm~\cite{Dokshitzer:1997in} in
$Z \to q\bar q$ (left) and $H \to gg$ (right) processes.
Focusing on the PG$_{\betaps=0}^\text{sdf}$ shower,
the plots show results with various subsets of ingredients.
A zero result indicates NNLL accuracy.
Only with 2-jet NLO matching~\cite{Hamilton:2023dwb}, double-soft
corrections~\cite{FerrarioRavasio:2023kyg},
$B_2$~\cite{Dasgupta:2021hbh,vanBeekveld:2023lsa} terms, 3-loop
running of $\as$~\cite{Tarasov:1980au,Larin:1993tp}, $K_2$ contributions~\cite{Banfi:2018mcq,Catani:2019rvy},
and the drift correction of this Letter does one obtain agreement
with the known NNLL
predictions~\cite{Banfi:2016zlc,2024FutureResummations}.
For this shower and observable, the drift correction dominates.

\begin{figure}
  \centering
  \includegraphics[width=\linewidth,page=1]{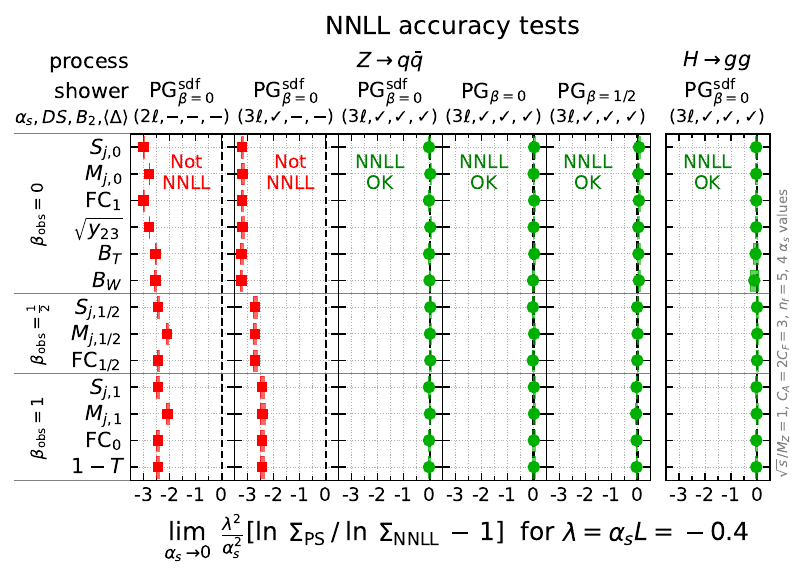}
  \caption{Summary of NNLL tests across observables and shower
    variants.
    Results consistent with zero (shown in green) are in agreement
    with NNLL.
    The observables correspond to the event shapes used in
    Ref.~\cite{Dasgupta:2020fwr} and they are grouped according
    to the power ($\beta_\text{obs}$) of their dependence on the
    emission angle.
    All showers that include the corrections of this Letter agree with
    NNLL.
  }
  \label{fig:summary-obs}
\end{figure}

Tests across a wider range of observables and shower variants are
shown in Fig.~\ref{fig:summary-obs} for a fixed value of
$\lambda = \as \ln v = -0.4$.
With the drifts and all other contributions included, there is good
agreement with the NNLL
predictions~\cite{
  Becher:2008cf,   
  Abbate:2010xh,   
  Chien:2010kc,    
  Monni:2011gb,    
  Becher:2012qc,   
  Hoang:2014wka,   
  Banfi:2014sua,   
  Banfi:2016zlc,   
  Banfi:2018mcq,   
  Bauer:2020npd,   
  2024FutureResummations%
}.

Earlier work on NLL accuracy had found that the coefficients of NLL
violations in common showers tended to be moderate for relatively
inclusive observables like event shapes~\cite{Dasgupta:2020fwr}.
In contrast, here we see that non-NNLL showers differ from NNLL
accuracy with coefficients of order one.
That suggests a potential non-negligible phenomenological effect.

\begin{figure}
  \centering
  \includegraphics[width=\linewidth,page=1]{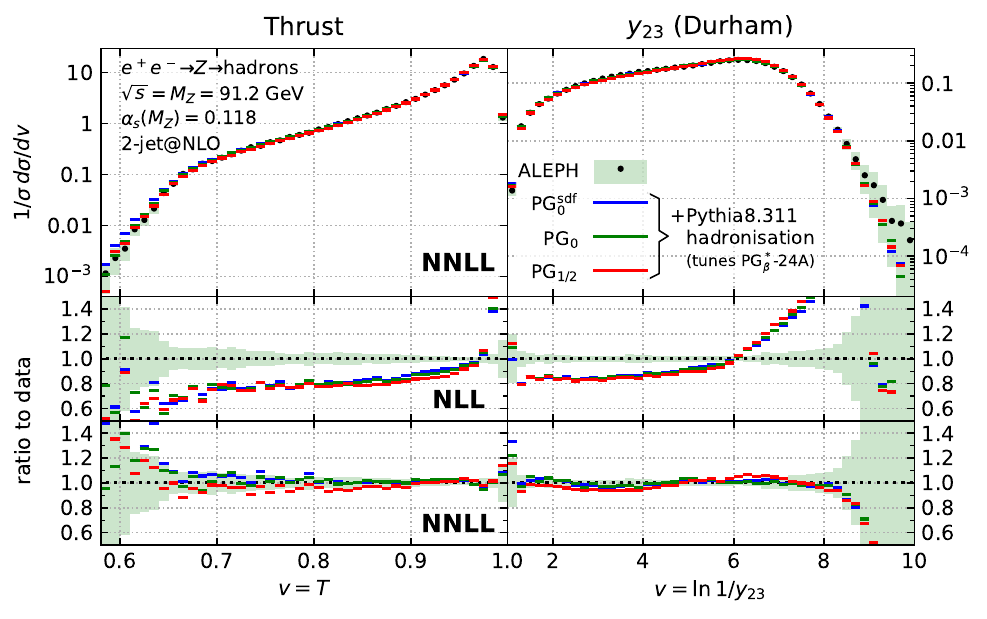}
  \caption{Results for the Thrust and Durham
    $y_{23}$~\cite{Catani:1991hj} observables with the PanGlobal
    showers compared to ALEPH data~\cite{ALEPH:2003obs}, using
    $\as(M_Z) = 0.118$.
    The lower (middle) panel shows the ratios of the NNLL (NLL) shower
    variants to data.
  }
  \label{fig:pheno-plot}
\end{figure}

Fig.~\ref{fig:pheno-plot} compares three PanGlobal showers
with ALEPH data~\cite{ALEPH:2003obs} using
Rivet~v3~\cite{Bierlich:2019rhm}, illustrating the showers in 
their NLL and NNLL
variants, with $\as^{\MSbar}(M_Z) = 0.118$ for both.
We use $2$-jet NLO matching~\cite{Hamilton:2023dwb}, and the NODS colour
scheme~\cite{Hamilton:2020rcu}, which guarantees
full-colour accuracy in terms up to NLL for global event shapes.
Our showers are implemented in a pre-release of
PanScales~\cite{vanBeekveld:2023ivn} v0.2.0, interfaced to Pythia
v8.311~\cite{Bierlich:2022pfr} for hadronisation, with
non-perturbative parameters tuned to
ALEPH~\cite{ALEPH:1996oqp,ALEPH:2003obs} and L3~\cite{L3:2004cdh} data
(starting from the Monash~13 tune~\cite{Skands:2014pea},
cf. Ref.~\cite{NNLLsupplement} \S\ref{sec:non-pert-tune}; the tune
has only a modest impact on the observables of
Fig.~\ref{fig:pheno-plot}).
The impact of the NNLL terms is significant and brings the showers into
good agreement with ALEPH data~\cite{ALEPH:2003obs}, both in terms of
normalisation and shape.
Some caution is required in interpreting the results: given that the
logarithms are not particularly large at LEP energies, NLO 3-jet
corrections (not included) may also play a significant role and should
be studied in future work.
Furthermore, the PanGlobal showers do not include finite quark-mass
effects.
Still, Fig.~\ref{fig:pheno-plot} suggests that NNLL terms have
the potential to resolve a long-standing issue in which a number of
dipole showers (including notably the Pythia~8 shower, but also the
PanGlobal NLL shower) required an anomalously large value of
$\as(m_Z) \gtrsim 0.130$~\cite{Skands:2014pea}  to achieve agreement with the
data.

The parton showers developed here are expected to achieve NNLL
(leading-colour) accuracy also for non-global event shapes such as
hemisphere or jet observables,
and $\as^n L^{n-1}$ (NSL)
accuracy~\cite{Frye:2016aiz,Kang:2018vgn,Kardos:2020gty,Bell:2020yzz,Anderle:2020mxj,Dasgupta:2022fim,vanBeekveld:2023lsa} 
for the 
soft-drop~\cite{Dasgupta:2013ihk,Larkoski:2014wba} family of 
observables, in the limit where
either their $z_\text{cut}$ parameter is taken small or
$\beta_{\text{\textsc{sd}}}>0$.
(We have not carried out corresponding logarithmic-accuracy tests,
because the small $z_\text{cut}$ limit renders them somewhat more
complicated than those of
Figs.~\ref{fig:log-dist-ratio}--\ref{fig:summary-obs}. In the case of
non-global event shapes, there exist no reference calculations.)
This is in addition to the NSL accuracy for energy-flow in a
slice~\cite{Banfi:2021owj,Banfi:2021xzn,Becher:2023vrh} and
$\as^n L^{2n-2}$ (NNDL) accuracy for subjet
multiplicities~\cite{Medves:2022ccw} that was already achieved with
the inclusion of double-soft
corrections~\cite{FerrarioRavasio:2023kyg}.

Next objectives in the programme of bringing higher logarithmic
accuracy to parton showers should include
incorporation of full triple-collinear splitting functions (as
relevant for experimentally important observables such as
fragmentation functions),
the extension to initial-state radiation,
and logarithmically consistent higher-order matching for a variety of
hadron-collider processes.
The results presented here, a significant advance in their own right,
also serve to give confidence in the feasibility and value of this
broad endeavour.

\begin{acknowledgments}
  We are grateful to Peter Skands for discussions and helpful
  suggestions on non-perturbative tunes in Pythia and to Silvia Zanoli
  for comments on the manuscript.
  This work has been funded by the European Research Council (ERC)
  under the European Union's Horizon 2020 research and innovation
  programme (grant agreement No.\ 788223, MD, KH, JH, GPS, GS) and under
  its Horizon Europe programme (grant agreement No.\ 101044599, PM),
  by a Royal Society Research Professorship
  (RP$\backslash$R1$\backslash$231001, GPS)
  and by the Science and Technology Facilities Council (STFC) under
  grants ST/T000864/1 (GPS), ST/X000761/1 (GPS), ST/T000856/1 (KH) and
  ST/X000516/1 (KH), ST/T001038/1 (MD) and ST/00077X/1 (MD).
  LS is supported by the Australian Research Council through a Discovery
  Early Career Researcher Award (project number DE230100867).
  BKE is supported by the Australian Research Council via Discovery Project DP220103512.
  We also thank each others' institutes for hospitality during the
  course of this work.
  Views and opinions expressed are those of the authors only and do
  not necessarily reflect those of the European Union or the European
  Research Council Executive Agency. Neither the European Union nor
  the granting authority can be held responsible for them.
\end{acknowledgments}

\bibliographystyle{apsrev4-2}
\bibliography{MC}

\newpage

\onecolumngrid
\newpage


\appendix

\makeatletter
\renewcommand\@biblabel[1]{[#1S]}
\makeatother

\section*{Supplemental material}

\subsection{Demonstration of relation between $\Delta K$ and drift}
\label{sec:div-theorem}

To help understand why Eq.~(\ref{eq:y-drift-constraint}) holds, we
choose here to focus on the ``non-Abelian'' or ``correlated-emission''
channels, which for a quark emitter correspond to the terms involving
$C_F n_f$ and $C_F C_A$ colour factors.
In general, in the soft limit, we define the correlated contribution
as being the part of the matrix element that remains after subtraction
of the double independent-emission contribution, which in the case of
a quark emitter corresponds to the $C_F^2$ component of the matrix
element.

The starting point is a definition for $\Delta K_1(y_\itilde)$, to be
understood with a renormalisation scale of $\mu = k_{t\itilde}$,
\begin{equation}
  \label{eq:nonAbK-defn}
  K_1 + \Delta K_1(y_\itilde)  
  =
  \frac{1}{\cB_\itilde} \left(
    \cV_{\itilde} + 
    \int d\Phi_{ij|\itilde}^\ps \, \cR_{ij}
  \right).
\end{equation}
It is convenient to define
\begin{equation}
  \label{eq:Rdiff}
  R(y_{\itilde}, y_d) =
  \frac{1}{\cB_\itilde}
 \left(
    \cV_{\itilde}\, \delta(y_\itilde - y_d)+ 
  \int
  d\Phi_{ij|\itilde}^\ps \, \cR_{ij}\, \delta(y_d(i,j) -
  y_d)
  \right),
\end{equation}
where $y_d(i,j)$ denotes the effective rapidity of the $i,j$
descendants, as produced by the shower.
In the correlated-emission channels, it is to be taken as the rapidity of
$i+j$.
The discussion can be adapted to the independent-emission channel by instead taking
$y_d$ to be the rapidity of either $i$ or $j$, choosing the one with
the larger shower ordering variable ($v_\ps = k_t e^{-\betaps |y|}$).

It is useful to introduce a few properties of $R(y_{\itilde}, y_d)$.
Firstly, a trivial rewriting of Eq.~(\ref{eq:nonAbK-defn}) is that
\begin{equation}
  \label{eq:nonAbK-defn-from-R}
  K_1 + \Delta K_1(y_\itilde)  
  =
  \int_{-\infty}^{+\infty} dy_d \, R(y_{\itilde}, y_d)\,.
\end{equation}
For the integral to converge, $R(y_{\itilde}, y_d)$ must vanish
sufficiently fast when $|y_d - y_\itilde|$ is large,
\begin{equation}
  \label{eq:R-large-deltay}
  R(y_{\itilde}, y_d) \to 0\,,\qquad \text{for } |y_d - y_\itilde| \gg 1\,,
\end{equation}
which can be seen as a consequence of the fact that for
large rapidity separations between $i$ and $j$, the shower reduces to
independent emission (cf.\ the PanScales
conditions~\cite{Dasgupta:2020fwr}).
A second property of $R(y_{\itilde}, y_d)$ is that
\begin{equation}
  \label{eq:KCMW-from-R}
  K_1 
  =
  \int_{-\infty}^{+\infty} dy_{\itilde} \, R(y_{\itilde}, y_d)\,,
\end{equation}
independently of $y_d$.
This corresponds to the statement that if one integrates over
$y_{\itilde}$, the shower will generate the correct rate of
soft-parton pairs for any $y_d$ --- it is an essential property of a
shower that has the correct double-soft contributions as in
Ref.~\cite{FerrarioRavasio:2023kyg}.
At large rapidities (but where the particles are still soft),
$R(y_{\itilde}, y_d)$ becomes a function of just $y_d -
y_{\itilde}$.
Denoting that function as $\bar{R}$, we have
\begin{subequations}
  \begin{align}
    \label{eq:R-to-Rbar}
    R(y_{\itilde}, y_d)  &\to \bar R(y_d - y_\itilde) ,\qquad \text{for } y_{\itilde},y_d \gg 1\,,
    \\
    R(y_{\itilde}, y_d)  &\to \bar R(y_\itilde - y_d) ,\qquad \text{for } -y_{\itilde},-y_d \gg 1\,.
  \end{align}
\end{subequations}
This then implies that $\Delta K_1(y_\itilde)$ tends to zero at large
$|y_\itilde|$, because Eqs.~(\ref{eq:nonAbK-defn-from-R}) and
(\ref{eq:KCMW-from-R}) become equivalent.

Now let us verify Eq.~(\ref{eq:y-drift-constraint}), which can be written as the
large-$Y$ limit of
\begin{align}
  \label{eq:Kint-two-integrals}
  \Delta K_1^\text{int}(Y) = \int_{-Y}^{+Y} dy_\itilde\, \Delta K_1(y_\itilde)
  &=
  \int_{-Y}^{+Y} dy_\itilde \int_{-\infty}^{+\infty} dy_d \,R(y_{\itilde}, y_d)
  - 
  \int_{-Y}^{+Y} dy_d \int_{-\infty}^{+\infty} dy_\itilde  \,R(y_{\itilde}, y_d)\,,
\end{align}
where, on the right-hand side, the last term is obtained by writing
$2Y K_1$ in terms of Eq.~(\ref{eq:KCMW-from-R}) for the factor of
$K_1$ and an integral over $y_d$ for the factor $2Y$.
Noting that the integration region where both rapidities are in the
range $-Y$ to $Y$ cancels between the two terms, we get
\begin{align}
  \frac12 \Delta K_1^\text{int}(Y)
  =
    \int_{-Y}^{+Y} dy_\itilde \int_{Y}^{+\infty} dy_d \,R(y_{\itilde}, y_d)
  - 
   \int_{-Y}^{+Y} dy_d  \int_{Y}^{+\infty} dy_\itilde \,R(y_{\itilde}, y_d)\,,
\end{align}
where the factor of $1/2$ accounts for the fact that we consider just
the positive infinite rapidity range (the negative infinite rapidity
range gives identical results, notably since
$R(y_\itilde,y_d) = R(-y_\itilde,-y_d)$).
Next, we make use of the property that the integrand is dominated by a
region where $y_\itilde$ and $y_d$ are not too different (cf.\
Eq.~(\ref{eq:R-large-deltay})), change variables to
$\Delta_y \equiv y_d - y_\itilde$ and $\bar y = y_\itilde - Y$, take
the limit $Y\to \infty$ and use Eq.~(\ref{eq:R-to-Rbar}) so as to write
\begin{equation}
  \label{eq:10}
  \frac12 \Delta K_1^\text{int}
  \equiv \lim_{Y \to \infty}
  \frac12 \Delta K_1^\text{int}(Y)
  =
   \int_{0}^{+\infty} d\Delta_y \int_{-\Delta_y}^{0} d\bar y \,\bar R(\Delta_y)
   -\int_{-\infty}^{0} d\Delta_y \int_{0}^{-\Delta_y} d\bar y \,\bar R(\Delta_y)\,.
\end{equation}
The $d \bar y$ integrations can be performed trivially and assembling
both terms, we obtain
\begin{equation}
  \label{eq:11}
 \frac12 \Delta K_1^\text{int}
  =
  \int_{-\infty}^{+\infty} d\Delta_y \Delta_y \,\bar R(\Delta_y)
  =
  \lim_{\itilde \to \text{\textsc{sc}}}
  \frac{1}{\cB_\itilde} \int
  d\Phi_{ij|\itilde}^\ps \, \cR_{ij}
  \times
  (y_d - y_\itilde)\,,
\end{equation}
which coincides with Eq.~(\ref{eq:y-drift-constraint}).

\subsection{Relation of $B_2$ coefficients with
  Refs.~\cite{Dasgupta:2021hbh,vanBeekveld:2023lsa}}\label{sec:B2-coeffs}

\logbook{b08e2cba40}{2023-07-21-NNLL-ee/maths/B2-games.nb}

\subsubsection{Overview of quark case}\label{sec:B2-quark-case}

Refs.~\cite{Dasgupta:2021hbh,vanBeekveld:2023lsa} consider a collinear
$q \to qg$ splitting probability, with original quark energy $E$,
splitting opening angle $\theta$ and outgoing quark energy $\zeta E$.
Eq.~(2.10) of Ref.~\cite{vanBeekveld:2023lsa} defines the splitting
probability as
\begin{equation}
  \label{eq:eq210-BDEHM}
  \frac{\as(\mu)}{2\pi}  \cP_q(\zeta,\theta)
  \equiv
  \frac{\as(\mu)}{2\pi} \left\{
  \frac{2C_F}{1-\zeta}\!\left[
    1 + \frac{\as(\mu)}{2\pi}\!\left(\!
      K_1 + \beta_0 \ln \frac{\mu^2}{(1-\zeta)^2 E^2 \theta^2}
      \right)
  \right]
  +
  \cB_1^q(\zeta)
  + \frac{\as(\mu)}{2\pi}\!\left(\!
    \cB_2^q(\zeta) + \cB_1^q(\zeta) \beta_0 \ln \frac{\mu^2}{E^2 \theta^2}
  \right)\!
  \right\}\!,
\end{equation}
where $\cB_1^q(\zeta) = -C_F(1+\zeta)$ is the finite part of the $q
\to qg$ splitting function and $\cB_2^q(\zeta)$ is to be found in
Refs.~\cite{Dasgupta:2021hbh,vanBeekveld:2023lsa}.
Specifically, we will need the integral of $\cB_2^q(\zeta)$, Eqs.~(3.6), (3.4) and (3.7)
of Ref.~\cite{vanBeekveld:2023lsa},
\begin{equation}
  \label{eq:B2int-nlo-result}
  \int_0^1 d\zeta \, \cB_2^q(\zeta) = -\gamma_q^{(2)} + \beta_0 X_{\theta^2}^q\,,
\end{equation}
with
\begin{equation}
  \label{eq:gammaq-X}
  \gamma_q^{(2)}
   = C_F^2\left(\frac{3}{8}-\frac{\pi^2}{2}+6\zeta_3\right)
    + C_FC_A\left(\frac{17}{24}+\frac{11\pi^2}{18}-3\zeta_3\right)
    - C_F n_f T_R\left(\frac{1}{6}+\frac{2\pi^2}{9}\right),\qquad
    X_{\theta^2}^q = C_F \left(\frac{2\pi^2}{3} - \frac{13}{2}\right).
\end{equation}
In that same collinear limit, up to order $\as^2$, let us write the parton shower's
splitting probability as
\begin{equation}
  \label{eq:Pgq-shower-parts}
  \frac{\aeff}{2\pi} P_{gq}(z)
  = 
  \frac{\as(k_t)}{2\pi} P_{gq}(z)
  \left[
    1 + \frac{\as(k_t)}{2\pi}\left(K_1 + B_2^\nlo(z) + B_2^\text{drift}(z)\right)
  \right],
\end{equation}
where we have separated $B_2(z)$ of Eq.~(\ref{eq:alpha-eff}) into two
pieces.
The piece $P_{gq}(z) B_2^\nlo(z)$ will integrate to $2C_F B_2^\text{int,\nlo}$,
while $P_{gq}(z) B_2^\text{drift}(z)$ will integrate to
$-2C_F \drift{\ln z}$ in Eq.~(\ref{eq:B2-relations-int}).
The fundamental constraint that we will impose is that the integrals
of Eqs.~(\ref{eq:eq210-BDEHM}) and (\ref{eq:Pgq-shower-parts}) should
yield identical results provided that they cover equivalent phase
space regions and include the full hard collinear domain.

Let us first examine the equivalence requirement if we consider limits
just on the $1\to 2$ phase space, with $z \equiv 1-\zeta$ (the
breaking of this equivalence at the level of the full $1\to 3$ phase
space will be examined below and included through
$B_2^\text{drift}(z)$).
In doing so, it is important to identify the relation between $k_t$
in the shower and the kinematics of a hard collinear splitting.
For the PanGlobal family of showers, it is straightforward to show
that
\begin{equation}
  k_t^\text{PG} \equiv z E \theta\,.
\end{equation}
Setting $\mu = (1-\zeta) E \theta$ in Eq.~(\ref{eq:eq210-BDEHM}), we can then write
\begin{equation}
  \label{eq:B2nlo-constraint}
  \lim_{\epsilon \to0}
  \int_\epsilon^{1-\epsilon} d\zeta
  \left[
    \frac{2C_F}{1-\zeta} K_1 + \cB_2^q(\zeta)
    + 2 \cB_1^q(\zeta) \beta_0 \ln (1-\zeta)
  \right]
  =
  \lim_{\epsilon \to0}
  \int_\epsilon^{1-\epsilon} dz P_{gq}(z) (K_1 + B_2^\nlo(z))\,,
\end{equation}
from which we deduce
\begin{equation}\label{eq:13}
  2 C_F B_2^\text{int,NLO}
  = \int_0^1 dz \left(\mathcal{B}^q_2(1-z)
  + \mathcal{B}^q_1(1-z)\left[2 \beta_0\ln z-K_1\right]\right)
  = -\gamma_q^{(2)} +
  C_F\beta_0\left(\frac{2\pi^2}{3}-3\right)+\frac{3}{2} C_F K_1\,.
\end{equation}
Recall that the $z$ dependence that we assign to $B_2^\nlo(z)$ is
arbitrary for event-shape NNLL accuracy, as long as the integral in
Eq.~(\ref{eq:DeltaKint-B2int-defs}) converges, i.e.\ $B_2^\nlo(z)$ has
to vanish sufficiently fast as $z \to 0$.
Our choice is
\begin{equation}
  \label{eq:B2nlo-z}
  B_2^\nlo(z) = \frac{3z}{2} B_2^\text{int,NLO}\,.
\end{equation}

\subsubsection{Overview of gluon case}\label{sec:B2-gluon-case}

The case of $B_2^\nlo(z)$ for gluon jets essentially follows the same
steps as that in the quark case modulo a few subtleties that require
extra care.
For the sake of conciseness, we only discuss the main steps in the
paragraphs below.

In the gluon case, the splitting probability is given by Eq.~(7.2) of
Ref.~\cite{vanBeekveld:2023lsa}.
The $z\leftrightarrow 1-z$ symmetry of gluon splittings, also
equivalent to the fact that parton showers generate gluon splitting by
summing over two dipoles, can be used to recast this equation as
\begin{align}
  \frac{\alpha_s(k_t)}{2\pi}\cP_g(\zeta,\theta)
  \equiv
  \frac{\alpha_s(k_t)}{2\pi}\left\{
  \frac{2C_A}{1-\zeta}\!\left(
    1 + \frac{\as(k_t)}{2\pi} K_1
  \right)
  +
  \cB_1^g(\zeta)
  + \frac{\as(k_t)}{2\pi}\left[
    \cB_2^g(\zeta) + 2 \cB_1^g(\zeta) \beta_0 \ln (1-\zeta)
  \right]\right\},
\end{align}
with $\mu=k_t\equiv (1-\zeta)\theta E$.
Here, we have explicitly summed the
$g\to gg$ and $g\to q\bar q$ channels by introducing
$\cB_1^g=\cB_1^{gg}+\cB_1^{qg}=C_A(\zeta(1-\zeta)-2)+n_fT_R(\zeta^2+(1-\zeta)^2)$
and
$\cB_2^g = \cB_2^{gg} + \cB_2^{qg}$ so that
\begin{equation}\label{eq:eq42-BDEHM}
  \int_0^1 d\zeta\, \cB_2^g(\zeta) = -\gamma_g^{(2)}+\beta_0X_{\theta^2}^g\,,
\end{equation}
with
\begin{equation}
  \gamma_g^{(2)}=\left(\frac{8}{3}+3\zeta_3\right)C_A^2-\frac{4}{3}C_An_fT_R-C_Fn_fT_R\,,\qquad
  X_{\theta^2}^g = -\left(\frac{67}{9}-\frac{2\pi^2}{3}\right)C_A+\frac{23}{9}n_fT_R\,.
\end{equation}
In writing this set of equations, we have performed a few
simplifications.
First, we have taken into account that, as in the quark case, the
coupling for PanGlobal is evaluated at the scale
$\mu=k_t\equiv k_t^\text{PG} = z E \theta$.
Then, Eq.~\eqref{eq:eq42-BDEHM} differs from the corresponding
equation~(4.2) in~Ref.~\cite{vanBeekveld:2023lsa} in that it does not
include the $\mathcal{F}_\text{clust.}^{C_A^2}$ contribution.
This contribution was an artefact of the use of an
mMDT~\cite{Dasgupta:2013ihk} procedure to define $z$ and $\theta$ in
the $C_A^2$ channel, which does not apply in our case. We also refer
to the discussion in Appendix~E of Ref.~\cite{vanBeekveld:2023lsa} for
more details.

With this in hand, the procedure follows exactly what we did above in
the quark case, yielding 
%
%
\begin{equation}
  2 C_A B_2^\text{int,NLO}
  =  \int_0^1 dz \left(\mathcal{B}_2^g(1-z)
  + \mathcal{B}_1^g(1-z)\left[2 \beta_0\ln z-K_1\right]\right)
  = -\gamma_g^{(2)} + \left[\left(\frac{2\pi^2}{3}-\frac{67}{18}\right)
    C_A + \frac{10}{9}n_fT_R\right] 
  \beta_0+\beta_0 K_1\,.
\end{equation}
This contribution has to be distributed over the two dipoles that
contribute to the emission rate from a gluon. When implementing this
in a shower, we also have the freedom to distribute this contribution
over the $g\to gg$ and $g\to q\bar q$ channels as well as the freedom,
already present in the quark case, to choose any explicit $z$
dependence as long as the gluon-splitting analogue of
Eq.~(\ref{eq:DeltaKint-B2int-defs}) is satisfied. 
In practice, we choose to have a unique $B_{2,g}^\nlo(z)$ common to
both splitting channels and take
\begin{equation}
B_{2,g}^\nlo(z) = \frac{24 C_A}{15 C_A+2n_fT_R} z\, B_2^\text{int,NLO},
\end{equation}
where the prefactor accompanies the shower-specific
partitioning prescription for the splitting functions $P_{gg}(z) =
C_A\frac{1+(1-z)^3}{2z}$ and $P_{qg}(z)=n_f T_R(1-z)^2$, for each of
the two dipoles contributing to a given gluon.
Note that, for event shapes, we are allowed to take a unique
$B_{2,g}^\nlo(z)$ across $g \to gg$ and $g \to q\bar q$ splittings
since its contribution only appears in the Sudakov exponent where it
is summed over flavour channels.

\subsubsection{Drift contributions}\label{sec:b2-drfit}

Concentrating once again on the correlated-emission channels, let us
schematically write $B_2^\nlo(z)$ as
\begin{equation}
  \label{eq:B2nlo-schematic-def}
  K_1 + B_2^\nlo(z) \equiv
    \frac{1}{\cB_{\itilde\ktilde}} \left(
    \cV_{\itilde\ktilde} + 
    \int d\Phi_{ijk|\itilde\ktilde}^\nlo \, \cR_{ijk}
  \right),
\end{equation}
where in the quark branching channel $\itilde$ is a gluon, $\ktilde$
is a quark, $z = E_\itilde/(E_\itilde + E_\ktilde)$ and we assume
the same scale choice $\mu=z E \theta$ as above.
This is the extension of Eq.~(\ref{eq:nonAbK-defn}) to the collinear
region noting that (a) in the soft-collinear region, $\Delta K_1(y)$ and
$B_2(z)$ both go to zero and (b) the ``Born'' starting point is a collinear
$\itilde\ktilde$ splitting, rather than just a soft $\itilde$
emission.
In the correlated-emission channels, the $d\Phi_{ijk|\itilde\ktilde}^\nlo$
phase space map is organised such that the energy fraction carried by
the $ij$ pair is equal to that carried by $\itilde$,
$E_i+E_j = E_\itilde$, where $\itilde$ is the emitted gluon (the
calculation also preserves $\theta_{ij,k} = \theta_{\itilde\ktilde}$).%
\footnote{In the independent-emission channel, we would instead have $E_i = E_\itilde$ and
$\theta_{i,jk} = \theta_{\itilde \ktilde}$.}

Let us now suppose that we had a parton shower with tree-level $1\to3$
splittings and 1-loop $1\to 2$ splittings.
That shower would implicitly involve a $B_2^\ps(z)$.
In what follows, we will work out the relation between the integral of
$B_2^\ps(z)$ and that of $B_2^\nlo(z)$ and observe that this
difference depends only on the behaviour of the shower in the soft
collinear region, where the triple-collinear structure reduces to the
double-soft structure.
This will be useful, because event-shape NNLL accuracy will be
sensitive just to the integral of $B_2(z)$, through its impact on
the Sudakov form factor.
Thus, even if our actual shower does not have the full triple-collinear
structure, the integral of $B_2^\ps(z)$ that we determine for an
imaginary shower that does have that structure will be sufficient to
achieve NNLL event-shape accuracy in the actual shower.\footnote{An
  analogous statement holds for the large-angle region, where global
  NNLL event-shape accuracy could have been obtained without the full
  large-angle double-soft structure, as long as the double-soft structure is still correct in
  the soft-collinear limit and the shower has the correct integral of
  $\Delta K_1(y)$. 
  We explicitly tested this by modifying $\Delta K_1(y)$ to be some
  function proportional to $e^{-|y|}$ with the constraint that its
  integral should be correct.
  This gave event shape results that were consistent with NNLL accuracy.
}

The $B_2^\ps(z)$ function can be expressed as
\begin{equation}
  \label{eq:B2ps-schematic-def}
  K_1 + B_2^\ps(z) \equiv
    \frac{1}{\cB_{\itilde\ktilde}} \left(
    \cV_{\itilde\ktilde} + 
    \int d\Phi_{ijk|\itilde\ktilde}^\ps \, \cR_{ijk}
  \right),
\end{equation}
where $d\Phi_{ijk|\itilde\ktilde}^\ps$ reflects the parton-shower
kinematic map and associated partitioning of phase space.
In particular the shower map will not in general have $E_i+E_j =
E_\itilde$ or $\theta_{ij,k} = \theta_{\itilde\ktilde}$.
Next, defining $B_2^\text{drift} = B_2^\ps(z) - B_2^\nlo(z)$ so as to
obtain Eq.~(\ref{eq:Pgq-shower-parts}), we have
\begin{equation}
  \label{eq:B2int-drift-defn}
  2C_F B_2^\text{int,drift}
  \equiv
  \int_0^1 dz P_{gq}(z) B_2^\text{drift}(z)
  =
  \lim_{\epsilon \to 0}
  \int_\epsilon^1 dz \, P_{gq}(z) \left[B_2^\ps(z) - B_2^\nlo(z)\right].
\end{equation}
It is now convenient to introduce a representation of the integrand of
Eq.~(\ref{eq:B2nlo-schematic-def}) that is differential in the energy
fraction $z_d$ of the descendants (e.g.\ $z_d =z_i+z_j$ in the non-Abelian channel)
\begin{equation}
  \label{eq:Rnlo}
  K_1 + B_2^\nlo(z) = \int dz_d R^\nlo(z,z_d)\,,
  \qquad
  R^\nlo(z_\itilde,z_d) =
  \frac{1}{\cB_{\itilde\ktilde}} \left(
    \cV_{\itilde\ktilde} \delta(z_d - z_\itilde) + 
    \int d\Phi_{ijk|\itilde\ktilde}^\nlo \, \cR_{ijk}
    \delta(z_d - z_d(i,j))
  \right)\,.
\end{equation}
For the NLO phase-space map discussed above  $R^\nlo(z,z_d)$
is equal to $(K_1+B_2^\nlo(z)) \delta(z-z_d)$.
One can also define an analogue for the parton shower, $R^\ps(z,z_d)$,
which will have a more complex structure.
We can rewrite Eq.~(\ref{eq:B2int-drift-defn}) as 
\begin{equation}
  2C_F B_2^\text{int,drift}
  \label{eq:drift-in-terms-of-R-step2}
  =
    \lim_{\epsilon \to 0}
    \int_0^1 dz_d
    \int_0^1 dz P_{gq}(z) \left[R^\ps(z,z_d) - R^\nlo(z,z_d)\right] \Theta(z-\epsilon)\,.
\end{equation}
Within our assumption that the shower has the correct triple collinear
(and double-collinear virtual) content,
the total amount of radiation at a given
$z_d$ will be reproduced by the parton shower, i.e.
\begin{equation}
  \int_0^1 dz P_{gq}(z) R^\ps(z,z_d) =  \int_0^1 dz P_{gq}(z) R^\nlo(z,z_d)\,.
\end{equation}
Furthermore, as in the case of section~\ref{sec:div-theorem}, any 
sensible shower that satisfies the PanScales conditions will have
the property that the integral over $R^\ps(z,z_d)$ is dominated by the region
of $z \sim z_d$.
For finite $z_d$, the $\Theta(z-\epsilon)$ term in
Eq.~(\ref{eq:drift-in-terms-of-R-step2}) is irrelevant.
Therefore if we place an upper limit on the $z_d$ integral,
$z_d \ll \epsilon_d$, with $\epsilon \ll \epsilon_d \ll 1$, the
result will be unchanged,
\begin{equation}
  2C_F B_2^\text{int,drift}
  \label{eq:drift-in-terms-of-R-step3}
  =
  \lim_{\substack{
      \epsilon_d \to0\\\epsilon/\epsilon_d \to 0}}
    \int_0^{\epsilon_d} dz_d
    \int_0^1 dz P_{gq}(z) \left[R^\ps(z,z_d) - R^\nlo(z,z_d)\right] \Theta(z-\epsilon)\,.
\end{equation}
Given that the integral now just involves a region of small $z$ and
$z_d$, i.e.\ we are in the soft-collinear limit,
we can use the property that aside from an overall $1/z_d$ factor,
$R^\ps(z,z_d)$ is a function just of the ratio of $z_d/z$
\begin{subequations}
\begin{align}
  R^\ps(z,z_d)  &\overset{z,z_d \ll 1}{\longrightarrow} \frac{1}{z_d} \bar R^\ps(z_d/z)\,,
  \\
  R^\nlo(z,z_d) &\overset{z,z_d \ll 1}{\longrightarrow} \frac{K_1}{z_d} \delta(z_d/z-1)\,.
\end{align}
\end{subequations}
In this limit, we can also use the soft limit of $P_{gq}(z)$ as
$z\sim z_d \ll 1$, allowing us to write
\begin{subequations}
\begin{align}
  2C_F B_2^\text{int,drift}
  &=
  2C_F
    \lim_{\substack{
    \epsilon_d \to0\\\epsilon/\epsilon_d \to 0}}
   \int_0^{\epsilon_d} \frac{dz_d}{z_d}
   \int_0^1 \frac{dz}{z}
   \left[\bar R^\ps(z_d/z) - K_1 \delta(z_d/z-1)\right] \Theta(z -
   \epsilon) \\
  &=
    2C_F
    \lim_{\substack{
    \epsilon_d \to0\\\epsilon/\epsilon_d \to 0}}
    \int_0^\infty \frac{d\xi}{\xi}
    \ln \frac{\epsilon_d}{\xi \epsilon}
   \left[\bar R^\ps(\xi) - K_1 \delta(\xi-1)\right]\,.
\end{align}
\end{subequations}
To get the second line, we have replaced $z\to z_d/\xi$, used the fact
that $R^\ps(\xi)$ vanishes for $\xi \to 0$ to replace the lower limit
of the $\xi$ integrand with $0$, and exchanged the order of $z_d$ and
$\xi$ integrations.

Next, exploiting the fact that $\int_0^\infty d\xi/\xi \bar R^\ps(\xi)
= K_1$, we can see that the $\ln \epsilon_d/\epsilon$ contribution vanishes,
leaving
\begin{equation}
  \label{eq:final-B2-drift-result}
  2C_F B_2^\text{int,drift} =
  2C_F \int_0^\infty \frac{d\xi}{\xi}
  \ln \frac{1}{\xi}
  \,
  \bar R^\ps(\xi)
    \equiv -2C_F \drift{\ln z}\,,
\end{equation}
which is the basis for Eq.~(\ref{eq:B2-relations-int}).

Eq.~(\ref{eq:Delta-lnz}) expresses the $\drift{\ln z}$ in terms of
$\drift{y}$ and $\drift{\ln m_t}$.
The relation between $\drift{\ln m_t}$ and $\drift{\ln k_t}$ is
given by
\begin{equation}
  \label{eq:mt-kt-drift-diff}
  \drift{\ln m_t} - \drift{\ln k_t} = \int d\Phi^\ps_{ij|\itilde}\,
  \cR(i,j) \,\frac{1}{2}\ln \frac{m_{t,ij}^2}{k_{t,ij}^2}\,.
\end{equation}
Note that aside from the phase space map, the integrand does not
depend on $\itilde$.
Accordingly, as long as the phase space map covers the full
double-soft region, which it does for any soft collinear $\itilde$, we
are free to replace the parton-shower phase space map with an analytic one that
preserves $k_t$ and rapidity.
One can then deduce the result for
Eq.~(\ref{eq:mt-kt-drift-diff}) from the corresponding expressions in
the literature, e.g.\ Eqs.~(3.7)--(3.12) of
Ref.~\cite{Banfi:2018mcq},
\begin{equation}
  \label{eq:mt-kt-drift-diff-res}
  \drift{\ln m_t} - \drift{\ln k_t} = -\beta_0\frac{\pi^2}{12}\,.
\end{equation}
We have also verified this numerically for each of our shower maps.
\logbook{48d724e67d}{See also 2023-07-21-NNLL-ee/maths/beta0-pisqsq12-checks.nb}

\subsection{Multiple emission contribution}
\label{sec:cF-derivation}

In our definition, the function $\cF$ in Eq.~(\ref{eq:Sigma-main})
accounts for the difference between applying a separate condition
$\Theta(V(k_\itilde) < v)$ on each primary emission $\itilde$, versus
applying a single condition $\Theta(V(\{k_i\}) < v)$ on the full set
of emissions after branching.
It starts at order $\as^2$ and its NNLL component (starting from
$\as^2 L$) differs from the corresponding $\delta \cF$ contribution
which appears at NNLL accuracy in ARES~\cite{Banfi:2018mcq}, in that
the latter also includes certain order $\as$ terms (cf.\ Eq.~(2.45) of
Ref.~\cite{Banfi:2018mcq}).
Let us start by defining a primary inclusive emission density
\begin{equation}
  \label{eq:single-emsn-defn}
  d\rho_k
  = 
  2 \frac{dk_t}{k_t}\,
  dz P_{gq}(z)\,
  \frac{\aeff}{2\pi}
  \Theta(z > k_t)\,,
\end{equation}
which allows us to write the Sudakov factor from
Eq.~(\ref{eq:Sigma-main}) as
\begin{equation}
  \label{eq:sudakov-defn}
  S(v) = \exp \left[
    -2\int d\rho_k\,
    M(k)
    \Theta(V(k) > v)
  \right].
\end{equation}
The factor of two in the Sudakov reflects the presence of two
hemispheres.
The matching factor $M(k)$ will be irrelevant in the remainder of this
section because it tends to $1$ for $k_t \to 0$, but we will need it
later in section~\ref{sec:anl-nnll}. 
Up to NLL, one way of writing $\cF$ is as
\begin{equation}
  \label{eq:F-in-resummation}
  \cF_\nll = \frac{S(\epsilon v)}{S(v)}
  \sum_{n=0}^\infty
  \frac{1}{n!}
  \prod_{\itilde=1}^n 
  \left[
    2\int d\rho_{k_\itilde}\,
    \Theta(V(k_\itilde) > \epsilon v)
  \right]
  \Theta(V(k_{\tilde 1}, \ldots, k_{\tilde n}) < v),
\end{equation}
where $S(\epsilon v)$ exponentiates the virtual corrections for no
emission down to scale $\epsilon v$, the denominator $S(v)$ simply
divides out the Sudakov factor already included in Eq.~(\ref{eq:Sigma-main}),
while the sum and product account for the real emission of any number
$n$ of primary particles $\itilde$ at scales above $\epsilon v$.
The final $\Theta$-function represents the constraint on the emissions
from the requirement for the event-shape observable to have a value
less than $v$.
The parameter $\epsilon$ is to be taken small, such that $\ln \epsilon$
is kept finite or, equivalently, $v\ll\epsilon\ll 1$, so that one does not need to
resum $\ln \epsilon$-enhanced terms, for example in the ratio
$S(\epsilon v)/S(v)$.

Now let us extend Eq.~(\ref{eq:F-in-resummation}) to NNLL.
In the various manipulations along the way, we will discard
contributions that are beyond NNLL (i.e.\ that are $\as^n L^{n-2}$ or
higher).
We start by writing
\begin{multline}
  \label{eq:cF-nnll-step1}
  \cF_\nnll =
  \frac{S(\epsilon v)}{S(v)}
  \sum_{n=0}^\infty
  \bigg\{
   \frac{1}{n!}     
   \prod_{\itilde=1}^n 
  \left[
    2\int d\rho_{k_\itilde}
    \left(u_\itilde +
      \frac{\as}{2\pi}
      \int d\Phi_{i_a i_b|\itilde} \, \frac{\cR_{i_a i_b}}{\cB_{\itilde}}
      \left[u_{i_a} u_{i_b} - u_\itilde \right]
    \right)
        \Theta(V(k_\itilde) > \epsilon v)
  \right]
  \times
   \\
   \times
    \Theta(V(\{k_1\}, \{k_2\} \ldots, \{k_n\}) < v)
  \bigg\},
\end{multline}
where the $u_\itilde$ notation, inspired by generating-functionals,
means that in the second line, $\{k_i\}$ is just $k_\itilde$, while $u_{i_a}u_{i_b}$
means that $\{k_i\}$ is to be understood as $k_{i_a},k_{i_b}$, i.e.
\begin{subequations}
  \label{eq:uu-action}
  \begin{align}
    u_{\itilde} \Theta(V(\left\{k_1\right\}, \dots,
    \left\{k_i\right\},\dots, \left\{k_n\right\}) < v)
    &
      \equiv \Theta(V(\left\{k_1\right\}, \dots, k_\itilde,
      \dots, \left\{k_n\right\}) < v)\,,
    \\
    \quad u_{i_a}u_{i_b} \Theta(V(\left\{k_1\right\}, \dots,
    \left\{k_i\right\},\dots, \left\{k_n\right\}) < v)
    &\equiv \Theta(V(\left\{k_1\right\}, \dots, k_{i_a}, k_{i_b},
      \dots, \left\{k_n\right\}) < v)\,. 
  \end{align}
\end{subequations}
Note that the $\itilde \to i_a i_b$ branching is unitary in the
soft-collinear region that dominates $\cF$.
The individual emission constraint
$\Theta(V(k_\itilde) > \epsilon v)$ preserves this unitarity, and so
one should similarly constrain just $V(k) > \epsilon
v$ in the definition of $S(\epsilon v)$ of Eq.~(\ref{eq:cF-nnll-step1}).

Next, we take advantage of freedom in how to define the unresolved
emissions (e.g.\ $V(k_\itilde) < \epsilon v$), as
long as we use consistent definitions of unresolved in the Sudakov and
real emission contributions.
In particular, we choose to place the resolution condition on final
($i_a,i_b$) particles rather than the (potentially) intermediate
$\itilde$ particle, which gives
\begin{multline}
  \label{eq:cF-nnll-step2}
  \cF_\nnll =
  \frac{\bar S(\epsilon v)}{S(v)}
  \sum_{n=0}^\infty
  \bigg\{
   \frac{1}{n!}     
   \prod_{\itilde=1}^n 
  \left[
    2\int d\rho_{k_\itilde}
    \left(u_\itilde +
      \frac{\as}{2\pi}
      \int d\Phi_{i_a i_b|\itilde} \, \frac{\cR_{i_a i_b}}{\cB_{\itilde}}
      \left[u_{i_a} u_{i_b} - u_\itilde \right]
    \right)
        \Theta(V(k_{i_{ab}}) > \epsilon v)
  \right]
  \times
   \\
   \times
    \Theta(V(\{k_1\}, \{k_2\} \ldots, \{k_n\}) < v)
  \bigg\}.
\end{multline}
Relative to Eq.~(\ref{eq:cF-nnll-step1}) there are two changes.
Firstly, at the end of the square bracket on the first line,
$\Theta(V(k_{\itilde}) > \epsilon v)$ has been replaced by
$\Theta(V(k_{i_{ab}}) > \epsilon v)$, where $k_{i_{ab}}$ is defined to
be $k_\itilde$ for the $u_\itilde$ contribution, while for the
$u_{i_a}u_{i_b}$ contribution it is equal to a massless momentum with
the same transverse components and rapidity as
$k_{i_a}+k_{i_b}$. 
The shower and resummation have different maps from $\itilde$ to
$i_a i_b$.
However, both maps have the property that if one integrates over all
possible $\itilde$, one obtains identical final distributions of
$i_a i_b$.
Since the $\Theta(V(k_{i_{ab}}) > \epsilon v)$ condition is a
constraint on those final momenta, not on the (sometimes) intermediate
$\itilde$ momenta, the sum and product in Eq.~(\ref{eq:cF-nnll-step2})
will be the same in the shower and in the resummation (at least up to
NNLL).
The second change to note in Eq.~(\ref{eq:cF-nnll-step2}) relative to
Eq.~(\ref{eq:cF-nnll-step1}) is that $S(\epsilon v)$ has been substituted
with $\bar S(\epsilon v)$, defined as
\begin{equation}
  \label{eq:sudakov-bar-defn}
  \bar S(\epsilon v) = \exp \left[
    -2\int d\rho_k
    \left(
      \Theta(V(k) > \epsilon v)
      +
      \frac{\as}{2\pi}
      \int d\Phi_{a b|k} \, \frac{\cR_{a b}}{\cB_{k}}
      \left[\Theta(V(k_{ab}) > \epsilon v) - \Theta(V(k) > \epsilon v) \right]
    \right)
  \right].
\end{equation}
As compared to the $S(\epsilon v)$ of Eq.~(\ref{eq:sudakov-defn}), the $\Theta(V(k) > \epsilon v)$ factor
has been replaced by conditions that are specified in terms of final
momenta, either $k$ when there was no secondary branching or $k_{ab}$
when there was a branching (with $k_{ab}$ defined in analogy to
$k_{i_{ab}}$ above).
This modification is necessary in order for the exponentiated virtual
corrections in $\bar S(\epsilon v)$ to exactly match the phase space
in the real sum and product of Eq.~(\ref{eq:cF-nnll-step2}), as required by
unitarity.

Since the sum-product contribution in Eq.~(\ref{eq:cF-nnll-step2}) is
the same between the parton shower and the resummation, it will cancel
in the ratio $\cF^\ps/\cF^\resum$, leaving us with
\begin{align}
  \label{eq:cF-ratios-as-Sbar-ratios}
  \frac{\cF^\ps}{\cF^\resum}
  =
  \frac{\bar S^\ps(\epsilon v)}{\bar S^\resum(\epsilon v)}
  &=
    \exp \left[
      -2\int d\rho_k
      \left(
      \frac{\as}{2\pi}
      \int d\Phi_{a b|k}^\ps \, \frac{\cR_{a b}}{\cB_{k}}
      [\Theta(V(k_{ab}) > \epsilon v) - \Theta(V(k) > \epsilon v)]
    \right)
    \right].
\end{align}
To reach that result, it is useful to note that in
$\bar S^\resum$, the map $d\Phi_{ab|k}^\resum$ has the property that
$V(k_{ab})$ and $V(k)$ are identical, giving
$\bar S^\resum(\epsilon v) = S(\epsilon v)$.
Our next step is to write $d\rho_k$ explicitly in the
soft-collinear limit as $4C_F dk_t/k_t dy \as(k_t)/2\pi$ and to use
$V(k) \propto k_t \exp[-\betaobs y]$.
We then perform the $k_t$ integration within the limits set by the
difference of $\Theta$-functions in
Eq.~(\ref{eq:cF-ratios-as-Sbar-ratios}).
In doing so, it is convenient to approximate the scale of the coupling
as $k_t \sim v e^{\betaobs y}$, which is legitimate up to and
including NNLL when focusing on the ratio of $\cF^\ps / \cF^\resum$.
We then have
\begin{subequations}
\begin{align}
  \label{eq:cF-ratios-step2}
  \frac{\cF^\ps}{\cF^\resum}
  &=
  \exp \left[
    -8 C_F \int_0^{\ln 1/v_\text{hc}}
    \!\!\!\!
    dy
    \int \frac{dk_t}{k_t} \delta\!\left(\ln \frac{k_t}{v e^{\betaobs y}} \right)
    \frac{\as^2(k_t)}{(2\pi)^2}
    \int d\Phi_{a b|k}^\ps \, \frac{\cR_{a b}}{\cB_{k}}
    \left(
      \ln \frac{k_{t,ab}}{k_t} - \betaobs(y_{ab} - y)
    \right)
  \right]
  \\
  & = 
  \label{eq:cF-ratios-final}
    1 - \frac{8C_F}{\betaobs} T_2(v,v_\text{hc})
    \left[
      \drift{\ln k_t} -
      \betaobs \drift{y}
    \right],
\end{align}
\end{subequations}
where we have ignored any $\epsilon$ dependence (since we ignore
$\ln \epsilon$ contributions) and in the second line we have
transformed the $dy$ integral into a $1/\betaobs dk_t/k_t$ integral
so as to obtain the $T_2(v,v_\text{hc})$ function.
The final result corresponds to Eq.~(\ref{eq:cF-ps-v-resum}).
Note that for $\betaobs=0$, the result is to be understood as the
$\betaobs \to0$ limit of Eq.~(\ref{eq:cF-ratios-final}), with
$1/\betaobs T_2(v,v_\text{hc}) \to (\frac{\as(v)}{2\pi})^2\ln 1/v$.

\subsection{Equivalence with NNLL resummation}
\label{sec:anl-nnll}

\subsubsection{Relation to the ARES approach}

In the previous sections, we have given analytic arguments showing (i)
how we can incorporate into the shower the hard-collinear
$\mathcal{B}_2(z)$ computed analytically in
Refs.~\cite{Dasgupta:2021hbh,vanBeekveld:2023lsa}, (ii) how the
various drift contributions emerge in the parton shower
Sudakov and ultimately compensate for the effects of the shower's
double-soft map close to the observable
boundary, cf.~Eq.~(\ref{eq:adding-it-together}).
In this section, we show that these ingredients, together with 3-loop
running coupling, the CMW scheme ($K_1$ and $K_2$) and full
double-soft matrix-element corrections are sufficient for the shower
to achieve NNLL accuracy, for which we take ARES~\cite{Banfi:2018mcq}
as our reference.\footnote{The discussion below is presented at full colour
  accuracy.
  Our shower double-soft corrections are currently implemented only at
  leading-colour accuracy, consequently the NNLL terms in the shower
  are also only leading-colour accurate.
  Were they to be upgraded to full colour, we would expect to achieve
  full-colour NNLL accuracy, at least for processes with two coloured
  Born legs.  }

Our starting point is the shower result, Eq.~(\ref{eq:Sigma-main}).
Since the $\drift{\ln k_t}$ and $\drift{y}$ contributions all cancel,
cf.\ Eq.~(\ref{eq:adding-it-together}) and
sections~\ref{sec:div-theorem}, \ref{sec:b2-drfit} and
\ref{sec:cF-derivation},
we will leave them out in the rest of this section, or equivalently
work as if they were zero.
We then write Eq.~(\ref{eq:Sigma-main}) using the shorthand
Eq.~(\ref{eq:sudakov-defn})
\begin{align}\label{eq:nnll-proof-start}
  \Sigma^\textsc{ps}_\textsc{nnll} (v) =
  \mathcal{F}^\textsc{ps}_\textsc{nnll}(v)\,
  S^\textsc{ps}(v)\,.
\end{align}
It is useful to write the Sudakov form factor $S^\ps$ separated
into terms of different logarithmic order
\begin{equation}
  \label{eq:S-breakdown}
  \ln S^\ps = (\ln S)_\LL + (\ln S)_\NLL + (\ln S)_\NNLL
  + \ldots
\end{equation}
We can write the individual orders as integrals either over $k_t$ and
$z$ (as in Eq.~(\ref{eq:Sigma-main}) and in typical resummations), or
as an integral over the shower ordering variable $v_\ps$ and $z$.
The latter gives slightly more complicated expressions, but helps
connect with the actual shower algorithm.
On a first pass, readers may wish to set $\betaps=0$, in which case
$v_\ps \equiv k_t$.
One can also check explicitly that any dependence on $\betaps$ cancels
separately at each logarithmic order.
We take an observable that in the soft or collinear limit behaves as
$V(k)=f(z)g(y)k_te^{-\betaobs y}$ with $f(z)\to 1$ for $z\to 0$ and
$g(y)\to 1$ for $y\to \infty$.
Using $z=k_t e^{y}$, we can write
\begin{equation}
  \label{eq:kt-and-V-defs}
  k_t=z^{\frac{\betaps}{1+\betaps}}v_\textsc{ps}^{\frac{1}{1+\betaps}},
  \qquad
  V(k)=f(z)\, g(y)\, V_{\text{sc}}(k)\,, \quad \text{ with }\quad
  V_{\text{sc}}(k) = z^\frac{\betaps-\betaobs}{1+\betaps}v_\ps^\frac{1+\betaobs}{1+\betaps}\,.
\end{equation}
At LL, $P_{gq}(z)$ can be approximated as $2C_F/z$, the 3-jet
matrix-element correction factor $M(k)$ can be ignored, and the
observable can be approximated by its soft-collinear limit, yielding
\begin{equation}\label{eq:SLL}
  (\ln S)_\LL
  \equiv
  -\frac{4}{1+\betaps}\int_0^1 \frac{dv_\ps}{v_\ps} \int_{v_\ps}^1 dz
  \frac{\alpha_s(z^\frac{\betaps}{1+\betaps}v_\ps^\frac{1}{1+\betaps})}{2\pi}\frac{2C_F}{z}\,
  \Theta(z^\frac{\betaps-\betaobs}{1+\betaps}v_\ps^\frac{1+\betaobs}{1+\betaps}>v)\,.
\end{equation}
Note that we choose to include $3$-loop running for the coupling even
in the LL contribution. Strictly speaking, this introduces subleading
contributions in $(\ln S)_\LL$, but it simplifies the expressions for
the higher-order contributions, without compromising any of the
arguments of this section.
At NLL, we have
\begin{align}
  (\ln S)_\NLL
  \equiv
  & -\frac{4K_1}{1+\betaps}\int_0^1 \frac{dv_\ps}{v_\ps} \int_{v_\ps}^1 dz
    \frac{\alpha_s^2(z^\frac{\betaps}{1+\betaps}v_\ps^\frac{1}{1+\betaps})}{(2\pi)^2}\frac{2C_F}{z}\,
    \Theta(z^\frac{\betaps-\betaobs}{1+\betaps}v_\ps^\frac{1+\betaobs}{1+\betaps}>v)
    \nonumber\\
  & -\frac{4 B_1}{1+\betaps}\int_0^1 \frac{dv_\textsc{ps}}{v_\textsc{ps}}
    \frac{\alpha_s(v_\textsc{ps}^\frac{1}{1+\betaps})}{2\pi}\,
    \Theta(v_\textsc{ps}^\frac{1+\betaobs}{1+\betaps}>v),
    \label{eq:SNLL}
\end{align}
where we have included the $K_1$ NLL contribution and introduced
$B_1 = \int_0^1 dz \mathcal{B}_1(1-z)$.
In the second line, we have substituted $z=1$ in the argument of $\as$
and the observable so as to have purely $\as^n L^n$ terms (aside from
the higher-loop running coupling contributions).
The independence of $(\ln S)_\LL$ and $(\ln S)_\NLL$ 
of $\betaps$ can be verified, e.g.\ by changing variables from $v_\ps$
to $v_\text{obs} \equiv V_\text{sc}(k)$.
 
The NNLL contributions can be written as
\begin{align}
(\ln S)_\NNLL\equiv
  &-\frac{4K_2^\text{resum}}{1+\betaps}\int_0^1 \frac{dv_\ps}{v_\ps} \int_{v_\ps}^1 dz
  \frac{\alpha_s^3(z^\frac{\betaps}{1+\betaps}v_\ps^\frac{1}{1+\betaps})}{(2\pi)^3}\frac{2C_F}{z}\,
  \Theta(z^\frac{\betaps-\betaobs}{1+\betaps}v_\ps^\frac{1+\betaobs}{1+\betaps}>v)
    \nonumber
  \\
  & -\frac{4 \tilde B_2^\ps}{1+\betaps}\int_0^1 \frac{dv_\textsc{ps}}{v_\textsc{ps}} 
    \frac{\alpha_s^2(v_\textsc{ps}^\frac{1}{1+\betaps})}{(2\pi)^2}\,
    \Theta(v_\textsc{ps}^\frac{1+\betaobs}{1+\betaps}>v)
    +
    H_1^\ps \frac{\alpha_s(Q)}{2\pi}
    +
    2 C_{1,\text{hc}}^\ps \frac{\alpha_s(v^{\frac{1}{1+\betaobs}})}{2\pi}
    +
    C_{1,\text{wa}}^\ps \frac{\alpha_s(v)}{2\pi}\,.
\end{align}
The $K_2^\text{resum}$ coefficient has
already been introduced, and we elaborate on 
the others below.
Note that each of the integrals is independent of $\betaps$, but some of the
coefficients multiplying them depend on $\betaps$.
Below, we will verify that this $\betaps$ dependence disappears
when summing over all contributions.

Let us start with $H_1^\ps$, which involves the $3$-jet matrix element
correction factor $M(k)$,
\begin{subequations}
  \begin{align}
    \label{eq:H1ps-def}
    H_1^\ps
    &=
      \lim_{\epsilon\to0} \left[
      \frac{4}{1+\betaps} \int_\epsilon^1 \frac{dv_\ps}{v_\ps}
      \int_{v_\ps}^1 dz
      P_{gq}(z) M(k) 
      - 
      \frac{4}{1+\betaps} \int_\epsilon^1 \frac{dv_\ps}{v_\ps}
      \int_{v_\ps}^1 dz
      P_{gq}(z)\right]
    \\
    &=
      \frac{4}{1+\betaps} \int_0^1 \frac{dv_\ps}{v_\ps}
      \int_{v_\ps}^1 dz
      P_{gq}(z) \left(M(k) - 1\right)\,.
  \end{align}
\end{subequations}
Note that $M(k)$ also accounts for the Jacobian associated with the
integration variables and exact phase space limits.
In the second line, the lower limit of the integration on $v_\ps$ could be 
taken to zero, since $M(k)\to 1$ when $k_t\to 0$.
Through unitarity, the shower is such that $\Sigma(v)$ is normalised
to one, i.e.\ $\Sigma(v=1)= 1$, which ensures that $H_1^\ps$ can be
determined fully from the real $3$-jet matrix element.
The result is independent of the choice of observable,
\begin{equation}
  \label{eq:H1-result}
    H_1^\ps = \frac{-C_F}{1+\betaps}
  \left(\frac{\pi^2}{3}(1+\betaps)+\frac{7}{2}(1-\betaps)\right).
\end{equation}
Next we examine $C_{1,\text{hc}}^\ps$,
\begin{subequations}\label{eq:C1hcps}
\begin{align}
  C_{1,\text{hc}}^\ps
  & = \frac{-2}{1+\betaobs} \int_0^1 dz P_{gq}(z)
    \log f(z)
    -\frac{2}{1+\betaobs} \int_0^1 dz \mathcal{B}_1(1-z)
    \left(\frac{\betaps-\betaobs}{1+\betaps} \log z\right)
  \\
  & = \frac{-2}{1+\betaobs} \int_0^1 dz P_{gq}(z)  \log f(z)
    +\frac{\betaobs-\betaps}{(1+\betaobs)(1+\betaps)} \frac{7C_F}{2}.
\end{align}
\end{subequations}
This has a contribution that accounts for the exact boundary of the
observable in the hard-collinear limit (the terms involving
$f(z)$) and another that accounts for the fact that in the second line
of Eq.~(\ref{eq:SNLL}) the boundary condition on $v_\ps$ was imposed
for $z=1$ rather than integrated over the actual $z$ dependence of
Eq.~(\ref{eq:kt-and-V-defs}).

One can proceed the same way for the wide-angle coefficient,
$C_{1,\text{wa}}^\ps$. In this limit, we can approximate
$P_{gq}(z)=2C_F/z$, and after changing variable from $z$ to $y$ one
gets
\begin{equation}\label{eq:C1waPS}
  C_{1,\text{wa}}^\ps
  = -8 C_F \int_0^\infty dy \log g(y).
\end{equation}
We now focus on the term of $(\ln S)_\NNLL$ proportional to
$B_\ps^{(2)}$.
This hard-collinear $\mathcal{O}(\alpha_s^2L)$ term receives
contributions from three sources: 
(i) $B_2(z)$ in Eq.~(\ref{eq:alpha-eff}),
(ii) a contribution involving $K_1\mathcal{B}_1(1-z)$, with
$\mathcal{B}_1$ coming from the product of the finite part of the
splitting function in Eq.~(\ref{eq:single-emsn-defn}) with $K_1$ from
Eq.~(\ref{eq:alpha-eff}),
and (iii) a leftover running-coupling correction from the NLL term.
The latter stems from the fact that the scale of the running coupling,
$k_t= z^{\betaps/(1+\betaps)} v^{1/(1+\betaps)}$, is evaluated at the
scale $v^{1/(1+\betaps)}$ in the NLL contribution $(\ln S)_\NLL$ given
above. Summing these contributions, we get
\begin{subequations}
  \begin{align}
    \tilde{B}_2^\ps
    & = \int_0^1 dz \left(P_{gq}(z) B_2^\ps(z)
      + K_1\mathcal{B}_1(1-z)
      - \mathcal{B}_1(1-z) \frac{2\beta_0\betaps}{1+\betaps}\log z\right)\\
    & = -\gamma_q^{(2)}
      - \beta_0 X^\ps
      + C_F \beta_0 \frac{\pi^2}{6}
      \,,
      \qquad \text{ with }\quad
      X^\ps=C_F\left(\frac{7}{2}\frac{\betaps}{1+\betaps}+3-\frac{2\pi^2}{3}\right).
  \end{align}  
\end{subequations}
In the second line, we have explicitly separated out a $\pi^2/6$ which
can be traced back to the $-\beta_0 \pi^2/12$ term in
Eq.~(\ref{eq:Delta-lnz})

In order to ease the comparison to the ARES formalism, it is helpful
to use Eq.~(\ref{eq:T-swap}) in order to rewrite the
$\beta_0 X^\ps$ contribution to $\tilde B_2^\ps$ in terms
of an extra contribution to $H_1^\ps$ and to $C_{1,\text{hc}}^\ps$.
This gives
\begin{align}\label{eq:SNNLL-v2}
(\ln S)_\NNLL\equiv
  &-\frac{4K_2^\text{resum}}{1+\betaps}\int_0^1 \frac{dv_\ps}{v_\ps} \int_{v_\ps}^1 dz
  \frac{\alpha_s^3(z^\frac{\betaps}{1+\betaps}v_\ps^\frac{1}{1+\betaps})}{(2\pi)^3}\frac{2C_F}{z}\,
  \Theta(z^\frac{\betaps-\betaobs}{1+\betaps}v_\ps^\frac{1+\betaobs}{1+\betaps}>v)
  \nonumber  \\
  & -\frac{4 \bar B^\ps_2}{1+\betaps}\int_0^1 \frac{dv_\textsc{ps}}{v_\textsc{ps}} 
    \frac{\alpha_s^2(v_\textsc{ps}^\frac{1}{1+\betaps})}{(2\pi)^2}\,
    \Theta(v_\textsc{ps}^\frac{1+\betaobs}{1+\betaps}>v)
    +
    \bar H_1^\ps \frac{\alpha_s(Q)}{2\pi}
    +
    2 \bar C_{1,\text{hc}}^\ps \frac{\alpha_s(v^{\frac{1}{1+\betaobs}})}{2\pi}
    +
    C_{1,\text{wa}}^\ps \frac{\alpha_s(v)}{2\pi}\,,
\end{align}
with
\begin{subequations}
\begin{align}
  \label{eq:B2psNNLL}
  \bar{B}_2^\ps
  & = \tilde{B}_2^\ps + \beta_0 X^\ps = -\gamma_q^{(2)} + C_F \beta_0 \frac{\pi^2}{6},\\
  \label{eq:H1psNNLL}
  \bar H_1^\ps
  & = H_1^\ps - 2 X^\ps = C_F\left(\pi^2-\frac{19}{2}\right), \\
  \label{eq:C1hcpsNNLL}
  \bar C_{1,\text{hc}}^\ps
  & = C_{1,\text{hc}}^\ps + X^\ps
    = \frac{-2}{1+\betaobs} \int_0^1 dz\, P_{gq}(z)  \log f(z)
  +C_F\left(3+\frac{7}{2}\frac{\betaobs}{1+\betaobs}-\frac{2\pi^2}{3}\right). 
\end{align}
\end{subequations}
One sees that this reorganisation explicitly removes the residual
dependence on $\betaps$.

We are now in a position to connect the above ingredients to the
corresponding ones in the ARES NNLL formalism where the cumulative
distribution can be recast as follows (neglecting N$^3$LL
corrections, as elsewhere in this section), 
\begin{align}\label{eq:sigma-resum}
  \Sigma_\textsc{nnll}(v)
  = e^{-R(v)}
  \left[
  1
  + \frac{\alpha_s(Q)}{2\pi}H^{(1)}
  + \frac{\alpha_s(v^{\frac{1}{1+\betaobs}})}{2\pi} 2 C^{(1)}_\text{hc}
  \right]\mathcal{F}_\textsc{NNLL}\,,
\end{align}
cf.\ Eq.~(2.45) of Ref.~\cite{Banfi:2018mcq}, which
can be viewed as a short-hand notation for Eq.~(2.39) of that same
reference.
At LL and NLL, changing variables from $v_\textsc{ps}$ to $k_t$ is
sufficient to see that Eqs.~(\ref{eq:SLL}) and~(\ref{eq:SNLL})
correspond to the ARES LL and NLL contributions to the Sudakov
radiator, and also capture the $3$-loop running-coupling contributions
to the NNLL Sudakov, as already discussed above.

The NNLL contributions in Eq.~\eqref{eq:SNNLL-v2} also map directly
onto a series of ingredients in ARES:
the $K_2^\text{resum}$ term in Eq.~\eqref{eq:SNNLL-v2} reproduces the corresponding
CMW in the ARES Sudakov;
the $\bar H_1^\ps$ is identical to the $H^{(1)}$ coefficient in
ARES (cf.\ Eq.~(2.44) of Ref.~\cite{Banfi:2018mcq});
the $\gamma_q^{(2)}$ part of $\bar{B}_2^\ps$ corresponds to the
one-loop $\gamma_\ell^{(1)}$ contribution to the hard-collinear
radiator, cf.\ section 3.2 of Ref.~\cite{Banfi:2018mcq}, noting the
different sign convention between $\gamma_q^{(2)}$ and
$\gamma_\ell^{(1)}$;
and the $C_F\beta_0\pi^2/6$ part of $\bar{B}_2^\ps$ reproduces the
$\delta g_3^{(\ell)}$ contribution to the ARES Sudakov given in
Eq.~(3.28).
Eq.~(3.7) therein further showcases the similar origin of
this contribution in the shower and ARES approaches.

Having dealt with the above shower contributions, we are only left
with those in $\mathcal{F}_\textsc{nnll}$, $\bar C_{1,\text{hc}}^\ps$
and $C_{1,\text{wa}}^\ps$.
For these, it is helpful to consider also Eq.~(2.39) of
Ref.~\cite{Banfi:2018mcq} and Eq.~(\ref{eq:cF-nnll-step2}) above.

Most of the structure is common thanks to the fact that the shower
reproduces the relevant real kinematic configurations, namely, a
single hard-collinear emission, or a single soft-wide-angle emission,
or a double-soft pair of emissions, each accompanied by an arbitrary
number of well-separated soft-collinear emissions.

The $\mathcal{F}_\textsc{nnll}$ in ARES also receives a contribution
from the 1-loop correction to the soft-collinear gluon emission, see
the second line of Eq.~(2.39) of Ref.~\cite{Banfi:2018mcq}, ultimately
contributing to $\delta F_\text{sc}$. This is implicitly present in
the shower formalism which produces this term, through
unitarity, by including a factor $1+\frac{\alpha_s(k_t)}{2\pi}K_1$ in
$\alpha_\text{eff}$.
In practice, this gives rise to the term proportional to
$K^{(1)}\equiv K_1$ in $R'_{\text{NNLL},\ell}$, cf.\ Eq.~(3.32) of
Ref.~\cite{Banfi:2018mcq}.

With these considerations in place, we are now only left with the
soft-wide-angle and hard-collinear contributions to
$\mathcal{F}_\textsc{nnll}$, and with the
$\bar C_{1,\text{hc}}^\ps$ and
$C_{1,\text{wa}}^\ps$ coefficients.
To address these, it is helpful to note that the Sudakov in
Eq.~(\ref{eq:sudakov-defn}) differs from the convention used in ARES
as the latter defines it using the observable computed in the
soft-collinear limit.
As already discussed in section~\ref{sec:cF-derivation}, as long as
this is done consistently in $S(\epsilon v)$ and in the phase-space
condition for resolved real emissions, cf.\ e.g.\
Eq.~(\ref{eq:cF-nnll-step2}), there is a degree of flexibility in
defining the Sudakov form factor, which also appears as a $1/S(v)$
prefactor in $\mathcal{F}_\textsc{nnll}$. This flexibility amounts to
reshuffling contributions between $S$ and $\mathcal{F}_\textsc{nnll}$.

Due to the way the soft-collinear approximation is used for the
observable in the ARES approach, the $C_{1,\text{wa}}^\ps$ term of our
Eq.~\eqref{eq:C1waPS} does not appear in the ARES Sudakov factor.
Instead, a corresponding contribution appears in the
$\delta \mathcal{F}_\text{wa}$ term of Eq.~(2.46) of
Ref.~\cite{Banfi:2018mcq}, as defined in Eqs.~(3.33)--(3.35) of
Ref.~\cite{Banfi:2014sua}.
The $f_\text{wa}(\eta,\phi)$ that appears there corresponds to our
$g(y)$.
It is easy to verify the exact equivalence for the specific example of
an additive observable, cf.\ Eq.~(C.23) of Ref.~\cite{Banfi:2014sua},
which gives
\begin{equation}
  \label{eq:ARES-Fwa}
  \frac{\as(Q)}{\pi} \delta \cF_\text{wa} =
  8C_F \frac{\as(v)}{2\pi} \cF_{\NLL} \int_0^\infty dy \,  g(y)\,,
\end{equation}
where on the right-hand side, we have translated to our notation.
Since our $C_{1,\text{wa}}^\ps$, as part of $S$, multiplies $\cF$ in
Eq.~(\ref{eq:nnll-proof-start}), the ARES and shower wide-angle
contributions are identical.
For more general observables, the additional contributions that start
at $\as^2$ are contained in $\cF$ both in ARES and in our analytic
analysis of the shower's prediction. 

A similar but slightly more involved argument holds in the
hard-collinear region.
There, the $\bar C_{1,\text{hc}}^\ps$ term corresponds to
two contributions in ARES: one from the hard-collinear constant
$C^{(1)}_{\text{hc},\ell}$ in Eq.~(2.41) of
Ref.~\cite{Banfi:2018mcq}, and a second one from the
$\mathcal{O}(\alpha_s)$ contribution, $\delta
\mathcal{F}^{(1)}_\text{rec}$, to $\delta
\mathcal{F}_\text{rec}$ in Eq.~(2.46).
The $C^{(1)}_{\text{hc},\ell}$ coefficient is obtained when imposing
the condition
\begin{equation}
  \label{eq:C1hc-condition}
  z^{-\betaobs} k_{t,\textsc{ares}}^{1+\betaobs}  >  v\,,
\end{equation}
where, for a collinear $\tilde q\to qg$ splitting,
$k_{t,\textsc{ares}} = z(1-z)\theta_{qg} E_{\tilde q} = (1-z)k_t$.
This is to be supplemented~\cite{Banfi:2014sua,Banfi:2018mcq} with a
recoil term $\delta\mathcal{F}_\text{rec}$, which accounts for the
ratio between Eq.~(\ref{eq:C1hc-condition}) and the actual observable
condition $f(z)k_t e^{-\betaobs y}>  v$.
Making use of $e^y = z/k_t$ and the relation between
$k_{t}^\text{\textsc{ares}}$ and $k_t$, one can rewrite the actual
observable condition as
\begin{equation}\label{eq:ktAres-obs-condition}
  f(z) z^{-\betaobs} \left(\frac{k_{t,\textsc{ares}}}{1-z}\right)^{1+\betaobs} > v\,,
\end{equation}
which allows us to determine
\begin{equation}
  \delta\mathcal{F}_{\text{rec}}^{(1)}
  = \frac{2}{1+\betaobs}\int_0^1 dz\, P_{gq}(z)
  \log\left(\frac{(1-z)^{1+\beta_\text{obs}}}{f(z)}\right).
\end{equation}
The sum of these two contributions gives
\begin{equation}
  C^{(1)}_{\text{hc},\ell} + \delta\mathcal{F}_{\text{rec}}^{(1)}
  =\frac{-2}{1+\betaobs} \int_0^1 dz\, P_{gq}(z)  \log f(z)
  +C_F\left(3+\frac{7}{2}\frac{\betaobs}{1+\betaobs}-\frac{2\pi^2}{3}\right)
  = \bar C_{1,\text{hc}}^\ps,
\end{equation}
reproducing Eq.~(\ref{eq:C1hcpsNNLL}).
Once this is satisfied, the ARES and shower hard-collinear
contributions only differ by a reshuffling similar to the one done at
large angle in Eq.~(\ref{eq:ARES-Fwa}), hence not affecting
NNLL accuracy.
Note that the critical connection between an integrated $B_2(z)$
($\beta_0 X$ type terms), $C^{(1)}_{\text{hc},\ell}$ and
$\delta \cF_\text{rec}$ has been commented on before in
Ref.~\cite{Anderle:2020mxj} in the context of NNLL calculations for
groomed jet observables.

With this, all the terms have now been mapped between our shower and
the ARES formalisms, guaranteeing that, with the new ingredients
introduced in this paper, our shower algorithms achieve NNLL accuracy.

\subsubsection{Expressions for 3-loop CMW running coupling}\label{sec:3-loop-cmw-running}

For practical implementation in our showers, we have implemented
Eq.~(\ref{eq:alpha-eff}) factorising the genuine CMW running coupling
from additional NNLL contributions, namely
\begin{equation}\label{eq:practical-aeff}
  \aeff(k_t) = \as^\textsc{cmw}(k_t) \times
  \left\{1+ \tanh\left[\frac{\as^{\MSbar}(k_t)}{2\pi}\left (\Delta K_1(y)+B_2(z) \right)
    + \left(\frac{\as^{\MSbar}(k_t)}{2\pi}\right)^2\Delta K_2\right]\right\}\,,
\end{equation}
with $\Delta K_2=K_2^\ps-K_2^\text{resum}$.
For the first factor, we use an expansion valid at NNLL
accuracy:
\begin{align}
  \as^\textsc{cmw}(k_t) =& 
  \frac{\alpha_s}{1+t}
  +\frac{\alpha_s^2}{(1+t)^2} \left(\frac{-b_1}{b_0}\ln(1+t)+\frac{K_1}{2\pi}\right)\\
  &+\frac{\alpha_s^3}{(1+t)^3}\left[
    \frac{-b_2}{b_0}t+\frac{b_1^2}{b_0^2}\left(t-\ln(1+t)+\ln^2(1+t)\right)
    + \frac{K_2^\text{resum}}{4\pi^2} - \frac{b_1 K_1}{b_0\pi} \ln(1+t)
    \right], \nonumber
\end{align}
with $\alpha_s\equiv\alpha_s^{\MSbar}(Q)$,
$t=2\alpha_s b_0\ln(k_t/Q)$, and the following coefficients for the
QCD $\beta$-function and CMW $K_2$ coefficient
\begin{subequations}
\begin{align}
  b_0 &= \frac{11 C_A-4 n_f T_R}{12\pi},\\
  b_1 &=\frac{17 C_A^2 - 10 C_A n_f T_R - 6 C_F n_f T_R}{24\pi^2},\\
  b_2 &= \frac{2857 C_A^3 + (54 C_F^2-615 C_FC_A-1415 C_A^2)2n_fT_R +
       (66 C_F + 79 C_A) 4 n_f^2 T_R^2}{3456 \pi^3},\\
  K_2^\text{resum} &= \left(\frac{245}{24}-\frac{67\pi^2}{54}+\frac{11}{6}\zeta_3+\frac{11 \pi^4}{180}\right) C_A^2 
        + \left(4\zeta_3-\frac{55}{12}\right) C_F n_f T_R
        + \left(\frac{10\pi^2}{27}-\frac{209}{54}-\frac{14\zeta_3}{3}\right) C_An_fT_R\nonumber\\
      &\phantom{=}\;- \frac{4}{27}n_f^2T_R^2 + \frac{1}{2}\pi b_0
        \left[\left(\frac{808}{27}-28 \zeta_3\right)C_A-\frac{224}{27} n_fT_R\right].
\end{align}
\end{subequations}

The factor in the curly brackets of Eq.~\eqref{eq:practical-aeff}
incorporates all the shower contributions beyond the CMW running.
The specific form we have used only introduces subleading
contributions and guarantees both that the correction is never
negative and that it can easily be treated as an acceptance factor in
the Sudakov veto algorithm of the shower.

\subsection{Non-perturbative tuning}\label{sec:non-pert-tune}

\label{sec:tuning}

\begin{table}
  \begin{tabular}{lr@.lr@.lr@.lr@.lr@.l}
    \toprule
    parameter
    & \multicolumn{2}{c}{PG$_0^\text{sdf}$-24A}\;
    & \multicolumn{2}{c}{PG$_0$-24A}\;
    & \multicolumn{2}{c}{PG$_{1/2}$-24A}\;
    & \multicolumn{2}{c}{PG$_0^\text{sdf}$-M13}\;
    & \multicolumn{2}{c}{Monash13}
    \\  \midrule
    $\alpha_s(M_Z)$                 & 0&118 & 0&118 & 0&118 & 0&118 & 0&1365 \\    
    use CMW for $\alpha_s$          & \multicolumn{2}{l}{true} & \multicolumn{2}{l}{true} & \multicolumn{2}{l}{true} & \multicolumn{2}{l}{true} & \multicolumn{2}{l}{false} \\    
    $n$ loops for $\alpha_s$        & \multicolumn{2}{l}{3} & \multicolumn{2}{l}{3} & \multicolumn{2}{l}{3} & \multicolumn{2}{l}{3} & \multicolumn{2}{l}{1} \\       
    $k_{t,\min}$ shower cutoff      & \multicolumn{2}{l}{$0.5\GeV$} & \multicolumn{2}{l}{$0.5\GeV$} & \multicolumn{2}{l}{$0.5\GeV$} & \multicolumn{2}{l}{$0.5\GeV$} & \multicolumn{2}{l}{$0.5\GeV$} \\
    \midrule
    {\tt StringPT:sigma}            & 0&3026 & 0&294  & 0&29   & 0&335 & 0&335\\
    {\tt StringPT:enhancedFraction} & 0&0084 & 0&0107 & 0&0196 & 0&01  & 0&01 \\
    {\tt StringPT:enhancedWidth}    & 1&6317 & 1&5583 & 2&0    & 2&0   & 2&0  \\
    {\tt StringZ:aLund}             & 0&6553 & 0&7586 & 0&6331 & 0&68  & 0&68 \\
    {\tt StringZ:bLund}             & 0&7324 & 0&7421 & 0&5611 & 0&98  & 0&98 \\
    {\tt StringZ:aExtraDiquark}     & 0&9713 & 0&7267 & 0&8707 & 0&97  & 0&97 \\    
    \bottomrule
  \end{tabular}
  \caption{
    Top rows: parameters for the QCD running coupling
    used in each of our showers, and the corresponding values in the
    Monash~13 tune for the Pythia~8 shower.
    Bottom rows: parameters used in the Pythia~8.311 hadronisation model when
    interfaced to each of our showers.
    Other non-perturbative parameters coincide with the Monash13 tune.
  }\label{tab:tunes}
\end{table}

\begin{figure}
  \includegraphics[width=0.25\textwidth,page=1]{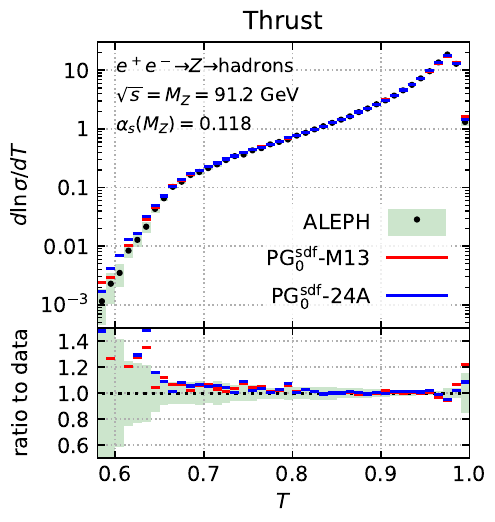}\hfill
  \includegraphics[width=0.25\textwidth,page=2]{plots-data/pg00sdf-v-lep.pdf}\hfill
  \includegraphics[width=0.25\textwidth,page=3]{plots-data/pg00sdf-v-lep.pdf}\hfill
  \includegraphics[width=0.25\textwidth,page=4]{plots-data/pg00sdf-v-lep.pdf}
  \\
  \includegraphics[width=0.25\textwidth,page=5]{plots-data/pg00sdf-v-lep.pdf}\hfill
  \includegraphics[width=0.25\textwidth,page=6]{plots-data/pg00sdf-v-lep.pdf}\hfill
  \includegraphics[width=0.25\textwidth,page=7]{plots-data/pg00sdf-v-lep.pdf}\hfill
  \includegraphics[width=0.25\textwidth,page=8]{plots-data/pg00sdf-v-lep.pdf}
  \\
  \includegraphics[width=0.25\textwidth,page=9 ]{plots-data/pg00sdf-v-lep.pdf}\hfill
  \includegraphics[width=0.25\textwidth,page=10]{plots-data/pg00sdf-v-lep.pdf}\hfill
  \includegraphics[width=0.25\textwidth,page=11]{plots-data/pg00sdf-v-lep.pdf}\hfill
  \includegraphics[width=0.25\textwidth,page=12]{plots-data/pg00sdf-v-lep.pdf}
  \\
  \includegraphics[width=0.25\textwidth,page=13]{plots-data/pg00sdf-v-lep.pdf}\hfill
  \includegraphics[width=0.25\textwidth,page=14]{plots-data/pg00sdf-v-lep.pdf}\hfill
  \includegraphics[width=0.25\textwidth,page=15]{plots-data/pg00sdf-v-lep.pdf}\hfill
  \includegraphics[width=0.25\textwidth,page=16]{plots-data/pg00sdf-v-lep.pdf}
  \caption{Results from the PG$^\text{sdf}_{\betaps=0}$ NNLL parton
    shower, supplemented with Pythia~8.311 hadronisation,
    compared to a range of
    data, using both the PG$_0^\text{sdf}$-24A and the
    PG$_0^\text{sdf}$-M13 tunes from
    Table~\ref{tab:tunes}.
    The two tunes have identical perturbative parameters, but
    different non-perturbative parameters and the plots illustrate
    that predictions for infrared safe observables (top three rows)
    are largely unaffected by the change in non-perturbative
    parameters, except at the edge of the perturbative region.
    }\label{fig:tuning}
\end{figure}

Table~\ref{tab:tunes} shows the non-perturbative parameters of
Pythia~8.311 that have been tuned relative to the default
Monash 2013 tune~\cite{Skands:2014pea}.
The tunes are based on ALEPH~\cite{ALEPH:1996oqp,ALEPH:2003obs} and
L3~\cite{L3:2004cdh} data and have been produced with our own 
proof-of-concept tuning framework.

In addition to modifying non-perturbative parameters, we also take 
$\as(M_Z) = 0.1180$ with a 3-loop running coupling.
In contrast, the Monash 2013 tune uses
$\as(M_Z) = 0.1365$ with a 1-loop running coupling.
Note that the latter does not include a $K_1$ contribution.
If one interprets $\as(M_Z) = 0.1365$ as a CMW-scheme coupling, the
corresponding $\MSbar$ value would be $\as^{\MSbar}(M_Z) = 0.1276$.
\logbook{}{$K=3.4540866883442094$, $0.12755 * ( 1+K*0.12755/(2*pi)) =
  0.136494$}

To appreciate the impact of the non-perturbative parameter choices,
Fig.~\ref{fig:tuning} shows results with the
PG$^\text{sdf}_{\betaps=0}$ shower and two tunes from
Table~\ref{tab:tunes}: the Monash13 tune (PG$_0^\text{sdf}$-M13, with
$\as(M_Z) = 0.1180$) and the dedicated tune presented here
(PG$_0^\text{sdf}$-24A).  (results for the other showers are broadly
similar).
For infrared safe observables (top three rows) the impact of the change in
parameters is negligible except (a) where experimental uncertainties grow
large and (b) deep in the 2-jet region where the Sudakov suppression
involves non-perturbative physics.
This gives confidence that the broad agreement that we see for these
observables has not simply been artificially engineered by the tuning
of the non-perturbative parameters.
For all observables in the top two rows and the Thrust Major on the third
row, our showers bring NNLL accuracy.
The remaining observables on the third row, i.e.\ the Thrust minor,
$y_{34}$ and $y_{45}$ have the property that they are non-zero
starting only from four or more particles and for these we do not
claim NNLL accuracy.\footnote{%
  For the Thrust minor and $y_{34}$ ($y_{45}$) the usual $\ln \Sigma$
  accuracy classification applies only if one requires three (four)
  hard jets.
  NNLL parton shower accuracy would then additionally require the
  shower to have 3-jet (4-jet) NLO accuracy.
}
Agreement remains generally good, though notably in the $4$ and
$5$-jet regions this is at least in part accidental, given the lack of
corresponding fixed-order matrix elements in the shower.
For infrared unsafe observables (bottom row), such as the distribution of the
number of charged tracks ($N_\text{ch}$), the particle momentum
distribution ($\xi_p = -\ln (2|\vec p|/Q)$) or the rapidity of
particles with respect to the Thrust axis ($y_T$), the tunes have a
significant impact.

We stress that this tuning exercise should be considered as
exploratory.
For example, we have not made any effort to address the question of
theory uncertainties.
From a perturbative viewpoint, we do not include any heavy-quark
effects (all quarks, including charm and bottom, are treated as
massless).
Also, we have made no efforts to tune the non-perturbative parameters
affecting the rates of various identified particles ($\pi^0$,
$\pi^\pm$, $K$, etc...).
Nevertheless, this tuning exercise does show that the good agreement
with infrared safe LEP $Z$-pole observables is not significantly
affected by variations of the non-perturbative parameters and that
distributions sensitive to non-perturbative physics improve after
tuning.



\end{document}